\documentclass[sigconf,edbt,hyphens]{acmart-edbt2021}

\usepackage{amsmath,amsfonts}
\usepackage{booktabs} 
\usepackage{algorithmic}
\usepackage{graphicx}
\usepackage{textcomp}
\usepackage{xcolor}
\usepackage[T1]{fontenc}
\usepackage{cleveref}
\usepackage{url}
\usepackage{boxedminipage}

\usepackage{subcaption}
\graphicspath{{./figures/}}

\usepackage[cachedir=cache,frozencache=true]{minted}
\newcommand{\inlinesql}[1]{\texttt{#1}}
\newmintinline[inlinecpp]{cpp}{}

\newlength{\mintednumbersep}
\AtBeginDocument{%
  \sbox0{\tiny00}%
  \setlength\mintednumbersep{\parindent}%
  \addtolength\mintednumbersep{-\wd0}%
}

\usepackage{xspace}
\newcommand*{\eg}{e.g.,\@\xspace}
\newcommand*{\ie}{i.e.,\@\xspace}
\newcommand*{\cf}{cf.\@\xspace}

\def\ParHead{\vspace*{1mm}\noindent\bf}
\def\ParHeadSmall{\noindent\bf}
\def\MiniHead{\\ \noindent\it}

\definecolor{agggreen}{HTML}{4DAF4B}
\definecolor{aggblue}{HTML}{377DB8}

\definecolor{motblue}{HTML}{377eb8}
\definecolor{motgreen}{HTML}{4daf4a}
\definecolor{motred}{HTML}{e41a1c}
\definecolor{motorange}{HTML}{ff7f00}
\definecolor{motpurple}{HTML}{984ea3}

\usepackage{tabularx}
\def\BibTeX{{\rm B\kern-.05em{\sc i\kern-.025em b}\kern-.08em
    T\kern-.1667em\lower.7ex\hbox{E}\kern-.125emX}}

\setcopyright{rightsretained}

\acmDOI{}

\acmISBN{978-3-89318-084-4}

\acmConference[EDBT 2021]{24th International Conference on Extending Database Technology (EDBT)}{March 23-26, 2021}{Nicosia, Cyprus} 
\acmYear{2021}

\begin{document}

\title{GeoBlocks: A Query-Cache Accelerated Data Structure\\ for Spatial Aggregation over Polygons}

\author{Christian Winter \ \ Andreas Kipf$^{\,\star}$ \ \ Christoph Anneser}
\author{Eleni Tzirita Zacharatou$^{\,\diamond}$ \ \ Thomas Neumann \ \ Alfons Kemper}
\affiliation{Technische Universität München \ \ \ \ MIT CSAIL$^\star$ \ \ \ \  Technische Universität Berlin$^\diamond$}
\affiliation{\{winterch, anneser, neumann, kemper\}@in.tum.de  \ \ \ \ \
kipf@mit.edu  \ \ \ \ \ eleni.tziritazacharatou@tu-berlin.de}

\renewcommand{\shortauthors}{Winter et al.}

\begin{abstract}
As individual traffic and public transport in cities are changing, 
city authorities need to analyze urban geospatial data to improve transportation and infrastructure.
To that end, they highly rely on spatial aggregation queries that extract summarized information from point data (\eg Uber rides) contained in a given polygonal region (\eg a city neighborhood). 
To support such queries, current analysis tools either allow only predefined aggregates on predefined regions and are thus unsuitable for exploratory analyses,
or access the raw data to compute aggregate results on-the-fly, which severely limits the interactivity.
At the same time, existing pre-aggregation techniques are inadequate since they maintain aggregates over rectangular regions.
As a result, when applied over arbitrary polygonal regions, they induce an approximation error that cannot be bounded. 

In this paper, we introduce GeoBlocks, a novel pre-aggregating data structure that supports spatial aggregation over arbitrary polygons. 
GeoBlocks closely approximate polygons using a set of fine-grained grid cells and, in contrast to prior work, 
allow to bound the approximation error by adjusting the cell size.
Furthermore, GeoBlocks employ a trie-like cache that caches aggregate results of frequently queried regions, 
thereby dynamically adapting to the skew inherently present in query workloads and improving performance over time.
In summary, GeoBlocks outperform on-the-fly aggregation by up to three orders of magnitude, achieving the sub-second query latencies required for interactive exploratory analytics.
\end{abstract}

\maketitle

\section{Introduction}
\label{sec:introduction}
Nowadays, the amount of geospatial data collected in cities is increasing rapidly, thanks to the widespread use of mobility applications such as Uber~\cite{uber:movement}. %
To analyze this data and make data-driven decisions, city officials and planners often rely on visualization frameworks that allow users to visualize data of interest at different 
spatial and temporal resolutions~\cite{urbane, uber:movement, TNCstoday, urban:bike, citi:bike}.
To generate common visualizations, such as heatmaps, visual tools perform \emph{spatial aggregation queries} that partition the data over different polygonal-shaped regions and then compute summarized aggregate information for each region.
To support \emph{exploratory analyses}, visual tools must provide interactive response times as high latency reduces the rate at which users make observations, draw generalizations, and generate hypotheses~\cite{visual:interactivity}. 
However, the sheer size of the data combined with the complexity of spatial queries prohibit interactivity, which severely limits analyses.
As shown in~\cite{DBLP:journals/pvldb/PandeyKNK18}, current
tools operating over raw geospatial data cannot produce results fast enough for interactive analysis.

On the bright side, interactive analyses are often repetitive in nature.
Analysts, for example, typically run multiple aggregate queries for the same area (\eg the city center) in a sequence, changing only the aggregate function (\eg count, sum) or the data attribute over which the aggregation is performed. 
Furthermore, they often focus on certain geospatial regions during their analysis.
They might, for example, iteratively resize the boundary of the spatial region of interest, extracting an aggregate every time, or calculate aggregates for neighboring, potentially overlapping, regions.
Such analyses can greatly benefit from query-driven materialization approaches that store and reuse intermediate or even full query results.

Naturally, in classical OLAP settings, query-driven materialization and result recycling are widely used and well understood~\cite{nagel:recycling, phan:materialization, shukla:materialized, shim:caching}.
However, these methods do not address multi-dimensional spatial data.
While methods have also been proposed for spatio-temporal OLAP queries, such as nanocubes \cite{lins:nanocubes} and the aR-tree~\cite{DBLP:conf/ssd/PapadiasKZT01, papadias:rtree}, these do not provide \textit{precision guarantees} for spatial aggregation queries over \textit{arbitrary polygons}. 
Both nanocubes and the aR-tree store aggregate information in a hierarchy of rectangles, maintained using a quadtree and an R-tree, respectively.
Therefore, they are designed for aggregate queries over rectangular regions while their precision depends on the granularity of the underlying index structure.
Using them to compute aggregates over \textit{polygonal} regions introduces an approximation error, which \textit{cannot be bounded}.
There are also some analysis tools, such as Uber Movement~\cite{uber:movement}, that rely on pre-computation to provide exact results for spatial aggregations over polygons. 
However, they require the polygonal regions to be pre-defined at aggregation time.
This assumes a priori knowledge of the workload and is thus not applicable in \textit{exploratory} analyses, where the query polygons are chosen \textit{ad-hoc}.

\begin{figure}
    \centering
  \includegraphics[width=\linewidth]{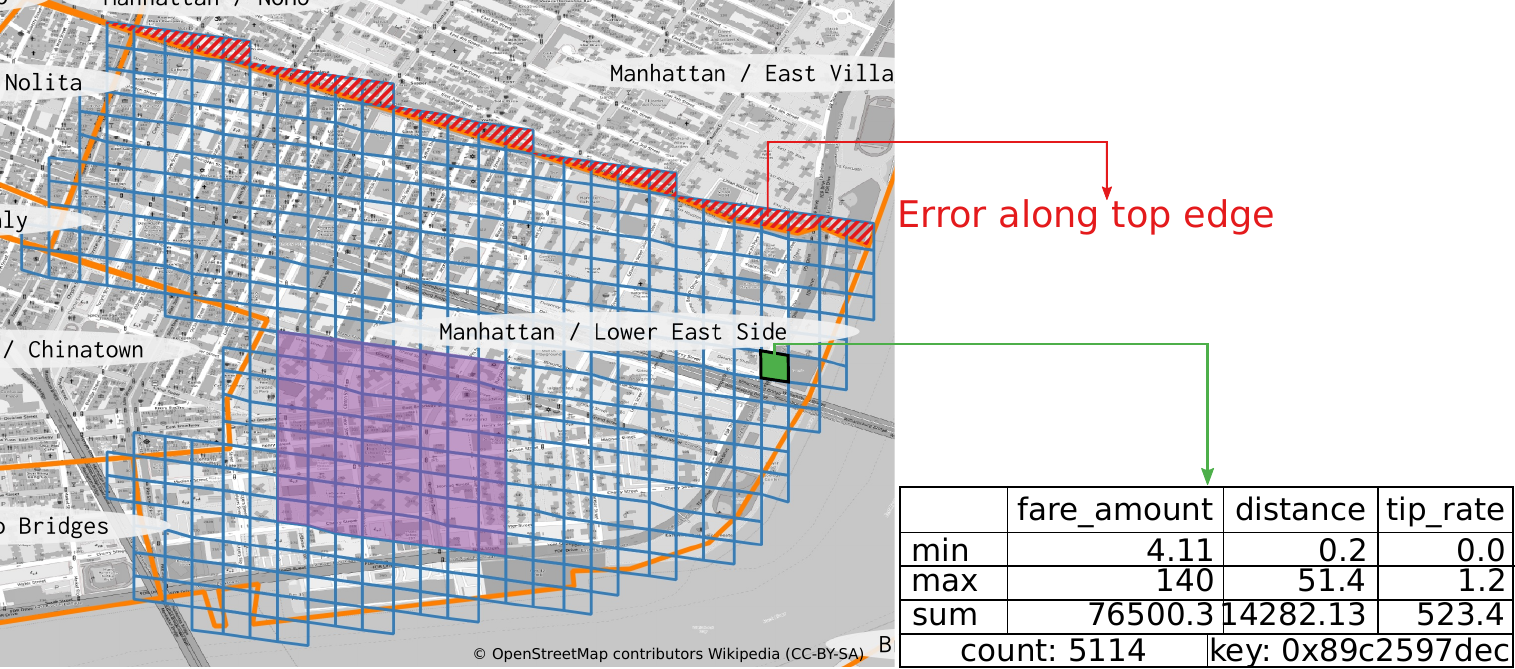}
  \caption{Cell covering (\textcolor{motblue}{blue}) of the Lower East Side (border in \textcolor{motorange}{orange}) with bounded error (\textcolor{motred}{red}), a cell aggregate (\textcolor{motgreen}{green}), and a cached commonly queried region (\textcolor{motpurple}{purple}).}
  \label{fig:geoblocks_motivational}

\end{figure}
We propose GeoBlocks, a novel pre-aggregating data structure for geospatial point data that guarantees error-boun\-ded results for spatial aggregation queries over arbitrarily shaped polygons.
Essentially, GeoBlocks are materialized views on geospatial point data that pre-compute filters and aggregations on pre-defined columns.
Instead of pre-computing aggregates over a hierarchy of rectangles as in prior work, GeoBlocks pre-compute aggregates over \textit{fine-grained grid cells}.
As depicted in Figure~\ref{fig:geoblocks_motivational}, GeoBlocks subdivide the spatial domain into grid cells, keeping aggregates for each individual cell.
We allow the user to specify the geospatial granularity, and thereby \textit{bound} the spatial approximation error.
In addition, we propose a trie-like data structure that caches aggregates for commonly queried regions in a compact manner, enabling even faster response times.
GeoBlocks are designed for historical point data and are thus write-once/read-only.
However, while GeoBlocks currently do not support updates, they can be adapted to do so, as we briefly discuss in Section~\ref{sec:discussion}. Our contributions are summarized as follows:
\begin{itemize}
    \item We propose GeoBlocks, the first, to the best of our knowledge, data structure that supports spatial aggregation over arbitrary polygons, while guaranteeing a bounded error.
    \item We develop a query-driven caching mechanism that further accelerates aggregate queries by leveraging the skew commonly found in exploratory query workloads.
\end{itemize}

The advantages of our approach are amply clear from our extensive experimental evaluation on real-world data.
The results show that GeoBlocks achieve up to three orders of magnitude speedup compared to on-the-fly aggregation approaches and support sub-second response times.

In the remainder of this paper, we first formalize the problem in \Cref{sec:problem}.
Section~\ref{sec:geoblocks} describes our approach, which we then experimentally evaluate in Section~\ref{sec:evaluation}. 
Section~\ref{sec:discussion} summarizes the key points discovered in the evaluation and discusses updates for GeoBlocks. 
Finally, we present an overview of related work in Section~\ref{sec:relatedwork} before concluding in Section~\ref{sec:conclusion}.

\begin{figure}
    \centering
  \includegraphics[width=\linewidth]{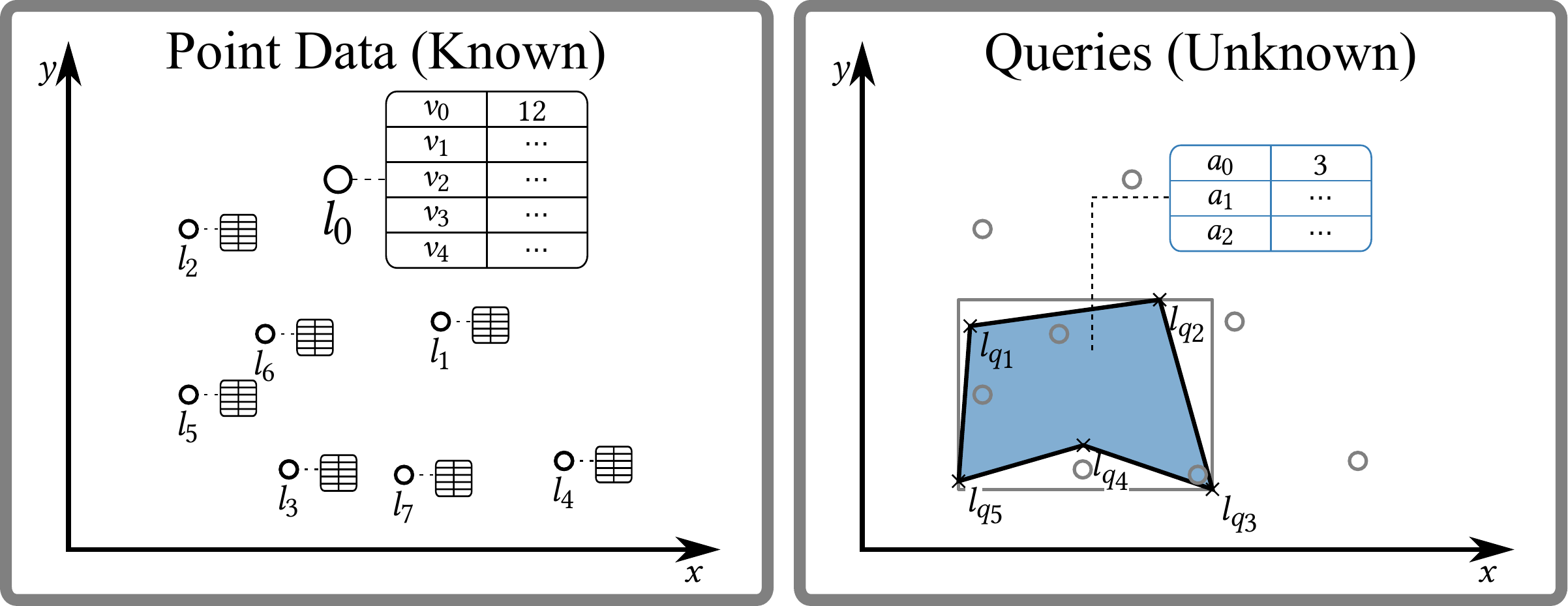}
  \caption{Problem overview: Calculating unknown aggregates $a$ from known points $P$ contained within an unknown query polygon $R$ (specified by its vertices $l_q$).}
  \label{fig:geoblocks_problem}

\end{figure}

\section{Problem Statement}
\label{sec:problem}
In this paper, we propose a new data structure to speed up the execution of spatial aggregation queries.
Formally, the query can be defined in SQL-like notation as follows:

\noindent 
\begin{boxedminipage}[b]{\linewidth}{}
\small
\texttt{SELECT AGG(P.$v_0$), \dots, AGG(P.$v_k$) FROM P \\
WHERE P.$l$ INSIDE R($l_{q1}, l_{q2}, \dots, l_{qm}$) [AND filterCondition]*}
\end{boxedminipage}

Given a set of annotated points of the form P($l$, $v_0, v_1, \dots, v_n$), where $l = (x_l, y_l)$ is the location of the point and $v_i$ are numerical or temporal attributes, this query extracts multiple aggregates $a_i = AGG(v_i)$ over all the points contained in a query region $R$.
The query region can be any \textit{arbitrary polygon}, and its geometry is defined by the locations of the polygon's vertices $l_{q1}, l_{q2}, \dots, l_{qm}$. 
The aggregates are non-holistic functions such as count, sum, min, max, or average.
Finally, the query can have zero or more \texttt{filterConditions} on the attributes. 

In exploratory interactive analyses, users can dynamically and unpredictably change not only the filtering conditions and the requested aggregates but also the polygonal query region.
The data points, on the other hand, are known a priori.
\Cref{fig:geoblocks_problem} presents an example of this scenario:
The left-hand side shows the input points that are located at ($l_0,\dots,l_7$) and have five attributes each. 
The right-hand side shows the query; the polygonal region is marked in blue, while three different aggregates are extracted.
As can be seen in the figure, this query applies the aggregation over the three points that are contained in the query region, located at $l_5$, $l_6$, and $l_7$.

Existing approaches for spatial aggregation queries, such as the aR-tree~\cite{papadias:rtree,DBLP:conf/ssd/PapadiasKZT01}, are designed for rectangular regions, and thus \textit{do not support arbitrary polygons}.
Applying them to the example of \Cref{fig:geoblocks_problem} requires to approximate the query polygon with a minimum bounding rectangle, displayed in grey, over which the aggregation is performed. 
This introduces an extra point in the results, $l_3$, which is outside the actual query region.  

\section{GeoBlocks}
\label{sec:geoblocks}

In this section, we first present the geospatial decomposition that forms the basis of our approach. 
We then discuss how we can quantify and \textit{bound} the error that this decomposition introduces. 
Next, we explain the core concepts of GeoBlocks, their storage layout, and the efficient evaluation of spatial aggregation queries using GeoBlocks. 
Finally, Section~\ref{subsec:geoblocks-querydriven} outlines our query-driven caching mechanism that further improves performance by leveraging the characteristics of the query workload. 
\Cref{tab:terms} provides an overview of the concepts introduced in this section.

\begin{table}[]
\begin{tabularx}{\linewidth}{@{}lX@{}}
\toprule
\textbf{cell}                & Rectangular area, hierarchically subdividable into four children   \\ \midrule
\textbf{cell level}          & Number of subdivisions performed on the spatial domain to obtain the cell \\ \midrule
\textbf{cell id/spatial key} & Unique one-dimensional identifier of a cell                             \\ \midrule
\textbf{block level}           & Level of grid cells in a GeoBlock \\ \midrule 
\textbf{cell aggregate}    & Aggregates of all tuples of a grid cell                                \\ \midrule
\textbf{cell covering}            & Error-bounded approximation of a polygon using cells                    \\ \bottomrule
\end{tabularx}
\caption{Terminology}
\label{tab:terms}
\end{table}

\subsection{Geospatial Decomposition}
\label{subsec:geoblocks-background}
\begin{figure}
    \centering
  \includegraphics[width=\linewidth]{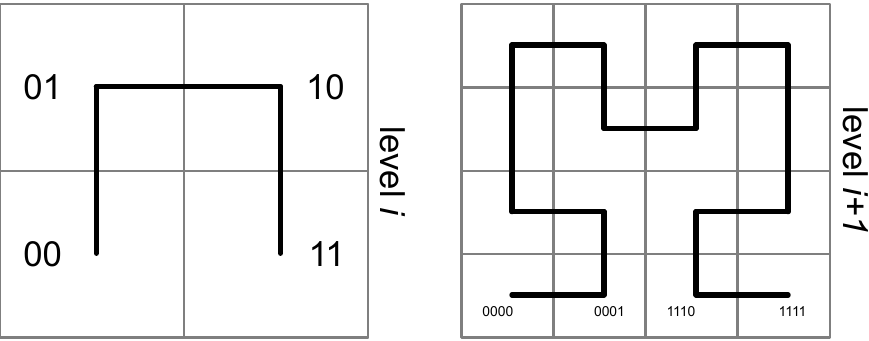}
  \caption{Hierarchical cell decomposition~\cite{kipf2020adaptive}.}
  \label{fig:geoblocks_cellenum}
\end{figure}

GeoBlocks rely on a hierarchical, quadtree-based spatial decomposition.
In this decomposition, a given area (cf. the outer rectangle in Figure~\ref{fig:geoblocks_cellenum}) is recursively subdivided into equally-sized smaller areas that we call cells. 
Each cell has four children, which leads to an exponentially growing number of $4^n$ cells after recursively subdividing a cell $n$ times.
We encode each subdivision using two bits, which allows us to uniquely identify a cell at level $n$ by concatenating the encoding of levels $0$ to $n$.
Equivalently, all cells at a given level can be enumerated using an \textit{order-preserving space-filling curve}.
Since children cells share a common prefix with their parent cell, containment tests are reduced to efficient bitwise operations. 
This encoding further allows storing cell ids in prefix-encoded index structures such as radix trees~\cite{kipf2020adaptive,DBLP:conf/icde/KipfLPPB0K18} or in learned indices~\cite{distance-bounded} to speed up containment queries.
Figure~\ref{fig:geoblocks_cellenum} shows the decomposition of a cell in four (level $i$) and 16 (level $i+1$) sub-cells, and the corresponding enumeration with a Hilbert curve.
Applying our decomposition strategy to the Earth's surface, we only need 64 bits to address every single square centimeter.
That way, we map two-dimensional geospatial locations (lat/long coordinates) to one-dimensional 64-bit keys. 
In our implementation, we use the Google S2 library~\cite{s2:general} to perform the spatial decomposition and cell enumeration. 
Note, however, that our approach is not restricted to S2 or the Hilbert curve.
Any other framework that supports recursive geospatial subdivisions and order-preserving cell enumerations can be used instead.

{\ParHeadSmall Point Approximation.} We map locations (\ie \textit{points}) to the smallest cell that contains them. 
The imprecision introduced by this approximation (\eg at most 6.1\,mm for any point in the US) is negligible, as the imprecision of GPS data is often orders of magnitude worse~\cite{van2015world}. 

{\ParHeadSmall Polygon Approximation.} Similarly, we approximate the query polygons on-the-fly by mapping them to a set of cells, possibly at different levels, as shown in \Cref{fig:coverings} (center and right).
We call this geometric approximation a \textit{cell covering}.
In our implementation, we calculate cell coverings using the S2 library. 

\subsection{Bounded Error}
\label{subsec:geo_error}
\begin{figure}%
	\centering
	 \includegraphics[width=\linewidth]{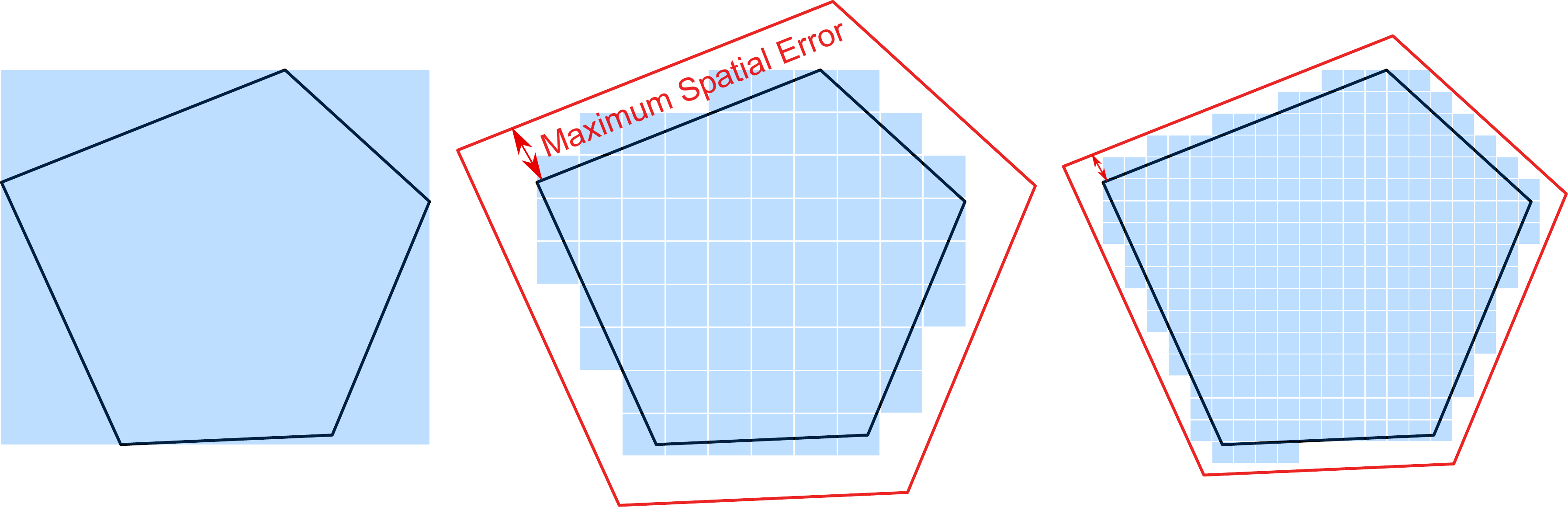}
    \caption{MBR (left) and two cell coverings with increasingly fine-grained resolution.}
	\label{fig:coverings}%
\end{figure}

Similarly to all geometric approximations, our cell covering introduces a spatial error. 
This is because all the cells that intersect the polygon outline, even minimally, are considered to be part of the polygon.
 However, in contrast to other coverings like the widely used minimum bounding rectangle (MBR), our cell covering is much more fine-grained.
As can be seen in \Cref{fig:coverings}, the cell covering approximates the polygon outline much more closely compared to the MBR. 
More importantly, the introduced approximation error can be bounded.
In fact, any point on the cell covering is within a distance $\sqrt{\epsilon_1^2 + \epsilon_2^2}$ from the polygon outline, where $\epsilon_1, \epsilon_2$ are the side lengths of the cell.
Clearly, the smaller the cell size, the smaller the approximation error.
Consequently, our cell covering can \textit{guarantee} a user-defined error bound, \ie a bound on the spatial distance between the approximate and the original polygon, by using an appropriately small cell size.
The MBR cannot guarantee such a bound, because its spatial extent, and thus its distance from the polygon outline, depends on the polygon's minimum and maximum coordinates in each dimension and cannot be controlled~\cite{distance-bounded}. 
The user can specify the error bound by choosing an appropriate cell level\footnote{From the table at \url{https://s2geometry.io/resources/s2cell_statistics}} so that the cell's diagonal is not greater than her desired error. This user-controlled and bounded spatial error is the only error in GeoBlocks. 
All further operations are exact and do not introduce any additional error. While the error bound should be the driving factor when selecting a cell level, there are other points to consider: (1) The cell diagonal is the maximum error, and the average error can be expected to be lower. (2) The cost of reducing the error is not linear. Per each level, the diagonal, and thereby the error bound, reduces by a factor of 2. At the same time, the number of grid cells, and thus the query input, grows by a factor of 4.

\subsection{Preprocessing}
\label{subsec:geo_preprocessing}
\begin{figure}
    \centering
    \includegraphics[width=\linewidth]{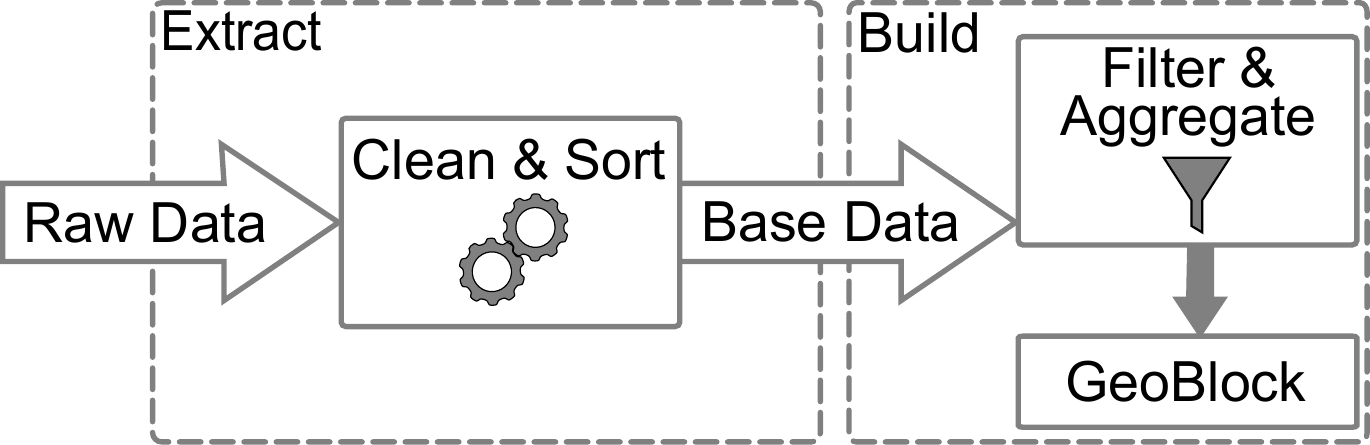}
    \caption{Creation of a GeoBlock in two phases. The extract phase is run once per dataset. The build phase is run for each filter and error bound combination.}
    \label{fig:process}
\end{figure}

In addition to transforming the two-dimensional input space to one-dimensional spatial keys, 
we perform some additional preprocessing steps on the known point data.
Our process, outlined in \Cref{fig:process}, consists of two phases, extract and build, and is similar to the ETL process traditionally applied in OLAP settings.  
In the first phase, we prepare the raw data by filtering outliers in the often dirty datasets and limiting the columns to those relevant and suitable for analysis. 
We furthermore sort the data by the generated one-dimensional spatial key. 
This extract phase is run exactly once per dataset and allows us to cheaply build GeoBlocks from the extracted base data.
The second phase, build, utilizes the clean and sorted base data to generate a GeoBlock in a single pass and thus in linear time.

{\ParHeadSmall Updates and Filters.} An important part of data analysis is filtering to gain insights into the desired subsets of the data. 
In our process, we could apply filters either before or after sorting the raw data. 
While the first option seems tempting, as it would reduce the number of tuples over which the expensive sorting has to be performed,  we decided to filter the data in the build phase. 
This way, we can utilize the sorted base data to quickly build GeoBlocks for different filter predicates, aggregates, and grid resolutions in a single pass.
Building new GeoBlocks quickly is especially useful in exploratory analyses, where the data and filters of interest might not be fully known a priori. 
However, the increased cost of sorting all data has to be amortized over multiple GeoBlocks and filter predicates. 
In reality, the sorting cost might be amortized immediately, as some exploratory queries might need to compare a subset of the data with the total.
Consider, for example, a query comparing the tip rate of expensive taxi rides (\inlinesql{WHERE fare\_amount > 20}) with that of all rides.
In this case, we would need to build a GeoBlock for all rides, and therefore sort the entire dataset anyway. 

Given $k$ different filter predicates with average selectivity $s$ and a total input size of $n$ tuples, we can calculate the runtime of building isolated GeoBlocks with filters before sorting, and incremental builds from sorted base data as follows:
\begin{equation}
k*(O(n)+O(sn*log(sn))+O(sn))
\label{eq:regular}
\end{equation}
\begin{equation}
O(n*log(n)) + k * O(n)
\label{eq:incremental}
\end{equation}
The isolated build (\ref{eq:regular}) has three phases, cleaning and filtering in $O(n)$, sorting in $O(sn*log(sn))$, and finally aggregating in $O(sn)$. Incremental builds (\ref{eq:incremental}) have a fixed component composed of cleaning and sorting in $O(n*log(n))$, followed by the incremental filtering and aggregation of the GeoBlock in $O(n)$.
For incremental builds to pay off, the sorting cost of the regular builds ($k*(O(sn*log(sn)))$ has to outweigh the initial cost of the incremental builds ($O(n*log(n))$).
As we only have runtime classes for each variant and de-facto runtimes will vary between systems and datasets, we cannot determine when amortization is reached solely depending on $k$ and $s$. 
However, we provide an in-depth experimental analysis of the amortization in \Cref{sec:evaluation}.

\subsection{Storage Layout}
\label{subsec:geoblocks-layout}

Once the filtering of the base data is completed, we can start aggregating and building a GeoBlock.
To build a GeoBlock, for each grid cell in the decomposed space, we compute a number of aggregates over all the tuples that it contains.
Empty cells that do not contain any tuples are omitted during aggregation as they would needlessly consume space. 
We refer to the aggregates of a grid cell as \textit{cell aggregates}.
A GeoBlock stores cell aggregates in ascending order of the cell's spatial key, which is the same sorting order as the one applied to the base data.
Moreover, a GeoBlock maintains a global header that combines all cell aggregates into a single GeoBlock-wide aggregate and contains additional metadata required for querying, such as the minimum and maximum cell id in the GeoBlock.

{\ParHeadSmall Cell Aggregate.}
Each cell aggregate stores pre-computed answers for spatial aggregation queries at the grid cell level. 
A cell aggregate consists of the cell's spatial key, the base data offset of the first tuple contained in the cell, and the number of contained tuples.
Furthermore, it maintains aggregates for all columns (both numeric and temporal attributes) in the extracted data. 
The maintained aggregates are the minimum, maximum, and sum of all values contained in the cell. 
Note that using the sum together with the tuple count allows us to also compute the average as sum/count.
Furthermore, the cell aggregate stores the minimum and maximum keys of the spatial column. 
The table in ~\Cref{fig:geoblocks_motivational} shows an example of a cell aggregate.

{\ParHeadSmall Aggregate Granularity.} As described in \Cref{subsec:geo_error},  the block level (\ie the granularity of the space decomposition) is defined by the user at build time.
However, it is also possible to adapt the granularity at a later time.
Building a more coarse-grained GeoBlock from an existing one is rather straightforward and does not require re-scanning the base data.
We can easily combine all cell aggregates of the finer-grained GeoBlock corresponding to a more coarse-grained grid cell in a single pass over the aggregates.
On the other hand, building a more fine-grained GeoBlock requires scanning and further subdividing the base data.

\subsection{Querying}
\label{subsec:geoblocks-query}
GeoBlocks support two variants of spatial aggregation queries. 
On the one hand, they support regular SQL \inlinesql{SELECT} queries that take a query polygon and produce a user-defined subset of the available aggregates. 
On the other hand, they support a specialized efficient implementation of \inlinesql{COUNT} queries that only report the number of points contained in a query polygon.
Such \inlinesql{COUNT} queries are commonly used in analytics, especially in the context of visualization.
\Cref{fig:geoblocks_motivational} shows an example query that extracts a set of aggregates over the Lower East Side region, which is approximated by a cell covering (marked in \textcolor{motblue}{blue}).
The answer is calculated by extracting and combining all the aggregates contained in the blue cells.

\begin{figure}
    \centering
    \includegraphics[width=\linewidth]{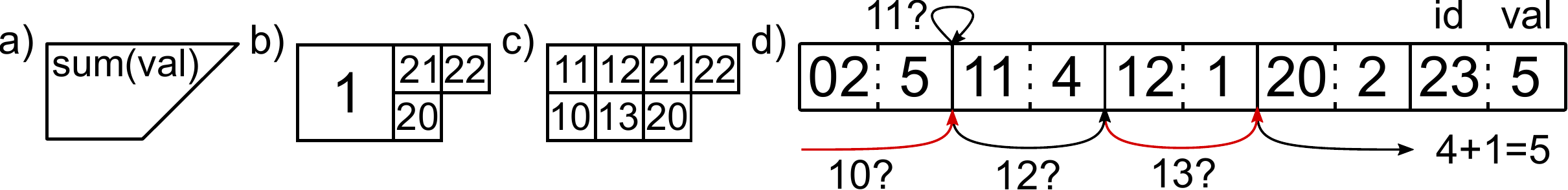}
    \caption{Query overview: Query polygon (a), cell covering (b), grid-cell representation of covering (c), and subquery for covering cell \textit{1} in the cell aggregates (d, \Cref{lst:geo_select} Line \ref{line:select_split} and following).}
    \label{fig:geo_query}
\end{figure}

To answer a spatial aggregation query over a polygonal region (\Cref{fig:geo_query}a), the polygon is approximated using a cell covering, as discussed in \Cref{subsec:geoblocks-background}. 
We compute a cell covering that conforms to the error bound (\Cref{fig:geo_query}b).
Note that the cell covering can have cells at different levels, and some of them might be larger than our grid cells.
Such larger cells can be easily mapped to smaller grid cells (\Cref{fig:geo_query}c) in the GeoBlock and offer further optimization potential, as discussed next.
The cell covering, however, cannot contain any cells smaller than the cells of the GeoBlock. 
Once we obtain the cell covering, we query the GeoBlock for each of the covering cells, as visualized for a \inlinesql{SELECT} query in \Cref{fig:geo_query}d.
We then combine these partial results to compute the final result for the entire query polygon.
In the following, we describe the query process for each cell of the covering. 
First, we use the GeoBlock's header to check if the cell overlaps with the GeoBlock at all. 
Thanks to the prefix-based containment checks, this is possible in constant time using the minimum and maximum cell id in the GeoBlock. 
Only if there is a possible overlap, we continue with the specific checks for \inlinesql{SELECT} and \inlinesql{COUNT} queries as follows:

\begin{listing}[t]
\begin{minted}[linenos,frame=lines,escapeinside=||,fontsize=\footnotesize,numbersep=\mintednumbersep,xleftmargin=4mm]{python}
lastAgg = 0
def selectQuery(polygon):
  queryCells = s2.coverPolygon(polygon)
  # Prune search range
  queryCells.pruneLess(globalHeader.minCell)|\label{line:sel_pres}|
  queryCells.pruneGreater(globalHeader.maxCell)|\label{line:sel_pree}|
  
  lastAgg = 0
  resultAggregates = initial
  for qcell in queryCells:
    # Map qCell to smaller childCells at the block level
    childCells = s2.childrenAtLvl(qcell, BLOCK_LVL) |\label{line:select_split}|
    for cell in childCells:
        getAggregates(cell, resultAggregates)
  return result
  
def getAggregates(cell, result):
  # Check the last results successor
  if lastAgg == 0:
    # Search initial header
    aggregate = allAggregates.upperBound(cell).prev|\label{line:select_searchs}|
    if aggregate.cell == cell:
      combineAggergates(aggregate, result)
    lastAgg = aggregate |\label{line:select_searche}|
  else:|\label{line:select_loops}|
    if lastAgg.next.cell == cell:
      lastAgg = lastAgg.next 
      combineAggregates(lastAgg, result) |\label{line:select_loope}|

      
\end{minted}
\caption{\inlinesql{SELECT} query}
\label{lst:geo_select}
\end{listing}

{ \ParHead \inlinesql{SELECT} Queries.}
\inlinesql{SELECT} queries have to look at all cell aggregates contained in the query cell.
Listing~\ref{lst:geo_select} presents the pseudo-code of the algorithm.
After a query cell has passed the first check, we try to further limit the search space to the overlapping area (Lines \ref{line:sel_pres} \& \ref{line:sel_pree}).
After splitting the query cell to smaller cells that match the GeoBlock's granularity if needed (Line \ref{line:select_split}), we locate the first intersecting grid cell using an upper-bound binary search (Lines \ref{line:select_searchs}~-~\ref{line:select_searche}). 
For all the following cells, we exploit the fact that cell aggregates are stored contiguously in ascending order. This allows us to iterate over the cell aggregates (Lines \ref{line:select_loops}~-~\ref{line:select_loope}) until we reach a grid cell not contained in the query cell, combining all cell aggregates along the way into the query result.

\begin{listing}[t]
\begin{minted}[linenos,frame=lines,escapeinside=||,fontsize=\footnotesize,numbersep=\mintednumbersep,xleftmargin=4mm, breaklines]{python}
def countQuery(polygon):
  queryCells = s2.coverPolygon(polygon)
  result = 0
  for c in queryCells:
    f_child = c.firstChildAtLvl(cell, BLOCK_LVL)
    l_child = c.lastChildAtLvl(cell, BLOCK_LVL)
    # Get first & last contained aggregate
    first = allAggregates.lowerBound(f_child) |\label{line:count_bs}|
    last = allAggregates.upperBound(l_child, first) |\label{line:count_be}|
    
    cnt = last.offset + last.count - first.offset |\label{line:count_res}|
    result += cnt
  return result
  
\end{minted}
\caption{\inlinesql{COUNT} query}
\label{lst:geo_count}
\end{listing}
{\ParHead \inlinesql{COUNT} Queries.}
Intuitively, we can answer \inlinesql{COUNT} queries faster than \inlinesql{SELECT} queries, as we can exploit the sorted order of the cell aggregates to calculate the count without accessing the cell aggregates of all grid cells that are contained in the query cell.
Specifically, \inlinesql{COUNT} queries can be answered using the count and offset values of only the first and the last cell aggregates that are contained in the query cell, as outlined in Listing~\ref{lst:geo_count}. 
Note that here we benefit from having larger query cells.
The larger the cells used in the covering, the fewer cell aggregates we need to access overall.
To find the first and last cell aggregates, we calculate the id of the first and last child of the query cell at our grid level. 
We then locate the first child in the aggregates using a lower bound binary search (Line \ref{line:count_bs}). 
Then, we use the position of the first child as a search start to locate the last child, again with a binary search (Line \ref{line:count_be}). 
Once we have located the aggregates of the first and last contained child, we can calculate (Line~\ref{line:count_res}) the resulting count in a range-sum manner as: 
\[
\mathrm{child_{last}}.\mathit{offset} + \mathrm{child_{last}}.\mathit{count} - \mathrm{child_{first}}.\mathit{offset}
\]

\subsection{Query-Cache Acceleration}
\label{subsec:geoblocks-querydriven}

While our cell aggregates can speedup queries significantly, there is further potential in pre-computing aggregates for frequently queried areas. 
This is based on the following key observations:
\begin{enumerate}
    \item Exploratory analyses are often repetitive in nature. Analysts, \eg may run consecutive queries for the same area to extract different aggregates (\ie using a different aggregate function, or aggregating over a different attribute).
    \item Furthermore, analysts might only iteratively change the shape or size of the query polygon. Consequently, part of the polygon's interior area remains unchanged. 
    \item Lastly, analytical queries often focus on a geographic subset of the whole data. For the analysis of the NYC taxi data, \eg the focus lies mostly on Manhattan, Brooklyn, and the airport regions, ignoring most suburbs~\cite{schneider:taxi}. 
\end{enumerate}
  
In all the above cases, it is reasonable to pre-aggregate small grid cells that are often queried together to avoid costly scans of individual cells. 
In our example in Figure~\ref{fig:geoblocks_motivational}, \eg we want to keep a single aggregate for the \textcolor{motpurple}{purple} region, instead of having to consult all 64 contained cell aggregates.

{\ParHead Determining Relevant Aggregates.} 
We want to determine the relevant areas that are worth being additionally pre-aggregated and cached, without making any assumptions about the expected query workload or the semantics of the indexed data. 
To achieve that, we use all previously seen queries as hints. 
Precisely, to determine whe\-ther an area is worth aggregating, we consider (i) the number of times it was queried, and (ii) its cell level.

For each query cell that intersects with the GeoBlock, we keep track of the number of times it was queried in a trie-like structure.
We then use these statistics to calculate cell scores. 
The score of a cell is the sum of the cell's hits and the hits of its parent. 
This score takes into account that child cells can be used to speed up queries for parent cells. 
We then sort all cells by descending score. 
When scores are identical, we sort by ascending level (coarser-grained cells come first).
As the last criterion, to ensure determinism, we sort by spatial key.
We chose the above metric as it is sufficient to properly and repeatably represent the skew in the experiments in our evaluation while being easy to understand and implement. 
However, we also identified some weaknesses of our metric: 

\begin{itemize}
    \item Smaller cells might overshadow slightly less frequently queried bigger cells. Consider, for example, the green and purple cells of Figure~\ref{fig:geoblocks_motivational} and assume that the green cell is queried just once more than the purple one. 
    Based on our metric, we would then aggregate the green cell even if the purple cell could have an up to $64\times$ bigger impact.
    \item The parent-child relationship is simplified: Children only cover parts of their parent but are treated as equally useful. 
    Furthermore, we do not consider calculating aggregates by combining the aggregates of the parent and siblings of a cell.
    For example, the count for a cell could be calculated by subtracting the count of its sibling cells from the count of its parent cell.
\end{itemize}

Our evaluation showed that these shortcomings have a minor impact, but we plan to investigate them further and address them, if needed, in our future work.

{\ParHead Aggregate Storage.}
\begin{figure}
    \centering
    \includegraphics[width=\linewidth]{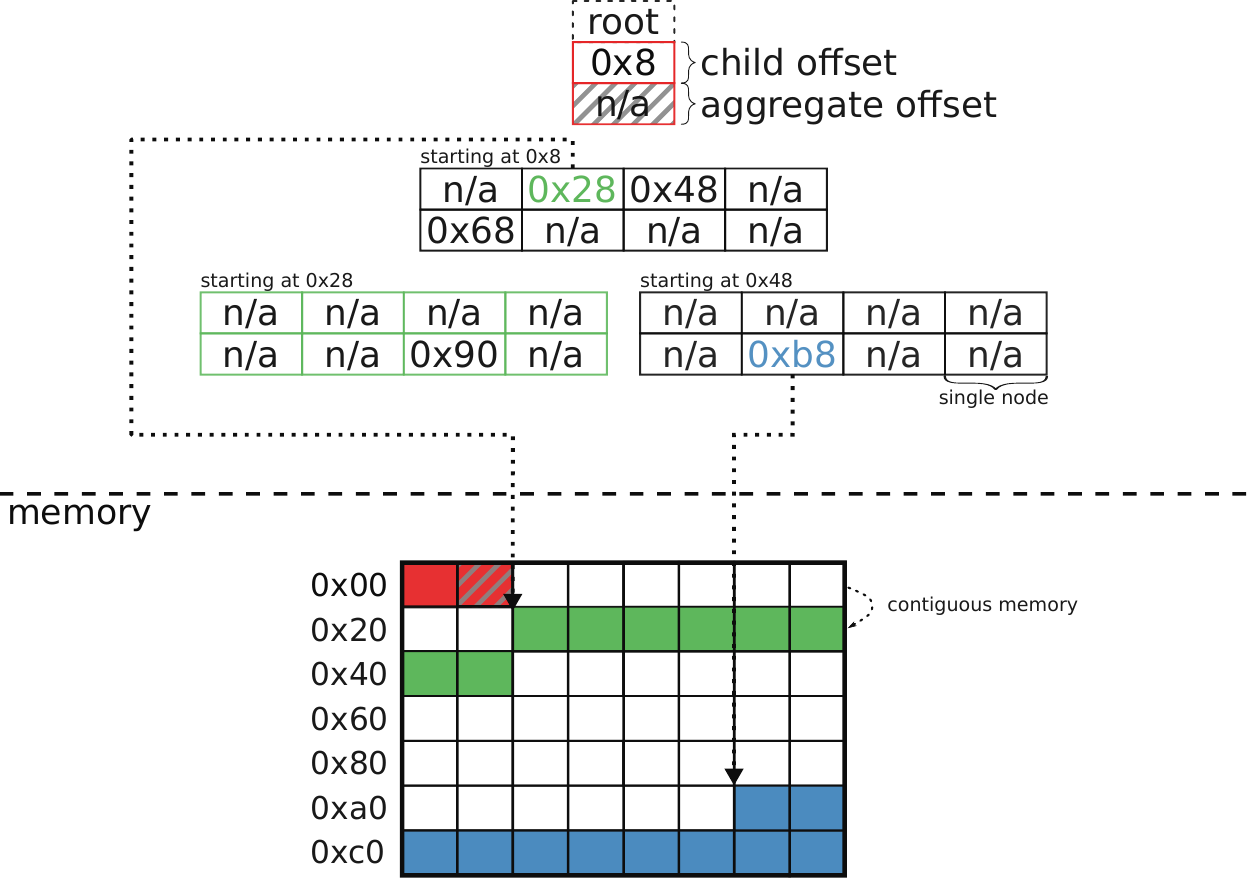}
    \caption{AggregateTrie with 40 byte aggregates and in-memory representation. Non-existent children (or aggregates) are marked with \textit{n/a} and are encoded as \texttt{0x0}.}
    \label{fig:geoblocks_aggtrie}
\end{figure}
We cache aggregates in a trie-like cache, which we call AggregateTrie.
Further, we allow the user to control the maximum size of the storage available for caching, and we store the AggregateTrie in-place with our cell aggregates and the filtered base data.
As the cells are strictly ordered, we can simply insert the most relevant unaggregated cell until the reserved area is filled. 
Figure~\ref{fig:geoblocks_aggtrie} shows an example AggregateTrie.

The storage for the aggregates is split into two parts. 
The first part (up until \texttt{0x90}) contains the trie structure, while the second part stores the actual aggregates.
The root of the trie corresponds to the cell level that can enclose our input data, which is typically just a small fraction of the possible earth-wide input space.
Each following trie-level encodes exactly one cell level, resulting in a fanout of 4. 
Since we store the AggregateTrie in-place, we chose a compact encoding storing all nodes contiguously. 
Nodes consist of just two 32-bit integers. 
The first one is the pointer to the first child in the AggregateTrie.
The second one is the pointer to the corresponding aggregate in the aggregate storage (\eg \textcolor{aggblue}{\texttt{0xb8}}). 
Pointers are encoded as 32-bit offsets from the start of the allocated memory region.
Both aggregates and nodes can be sufficiently encoded with an offset, as they are of fixed size.
Nodes occupy 8 bytes, while the size of the aggregates depends on the schema.
Since we store only the offset to the first child, we need to always allocate space for all children in a node, even for children that do not exist in the cache. 
This can be seen for the node starting at \textcolor{agggreen}{\texttt{0x28}}, where only one child has an aggregate and no other children or aggregates exist.
While this seems wasteful at first, the alternative would be to store four individual child offsets per node.
As children are only created and stored if they are needed, our encoding never occupies more storage than this alternative. 
In fact, our design is more space-efficient in all cases, except for this worst-case in the example above, where only one out of four children is required. 

\begin{figure}
    \centering
    \includegraphics[width=\linewidth]{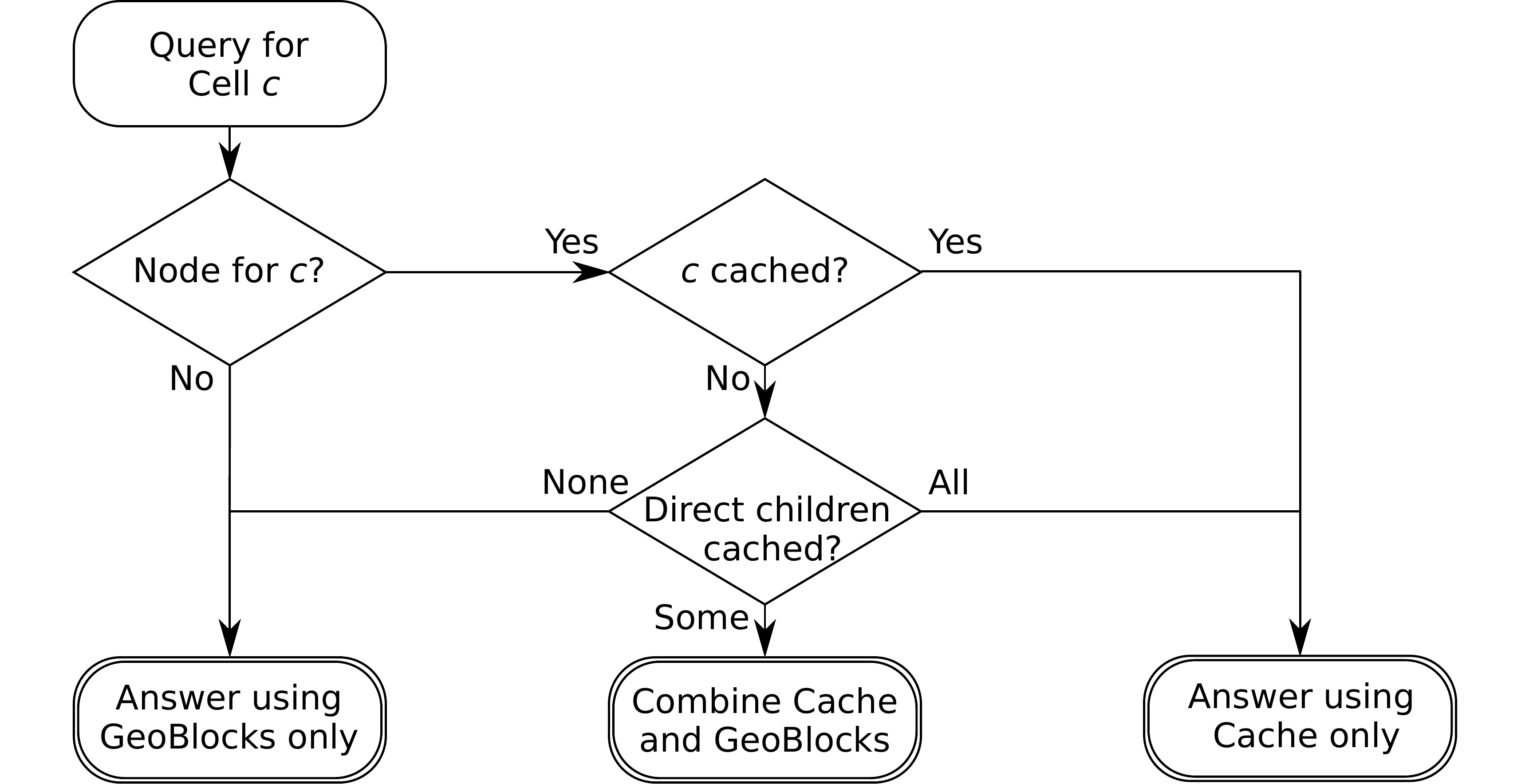}
    \caption{Overview of adapted query algorithm.}
    \label{fig:geoblocks_queryflow}
\end{figure}

{\ParHead Adapted Query Algorithm.}
We integrate the cached aggregates into the query algorithm (\cf Section~\ref{subsec:geoblocks-query}).
As the runtime of \inlinesql{COUNT} queries is mostly independent of the cell level since only the first and last grid cells are relevant, we do not expect noticeable speedups for them. Therefore, the adapted process, highlighted in Figure~\ref{fig:geoblocks_queryflow}, is only used for \inlinesql{SELECT} queries.

Once the pre-query checks are completed, we first probe the query cache and resort to the old algorithm only when necessary. 
For each query cell, we traverse the AggregateTrie to locate the corresponding node. 
If there is no node for this cell, we abort probing and answer the query with the old algorithm. 
Once the node corresponding to the cell is reached, there are two possible ways forward. If the cell is cached, \ie if it has a valid aggregate offset, the aggregate is extracted as a result. 
If the cell is not cached, there has to be at least one child at any level residing in our cache, as nodes are only created on demand.
While, theoretically, all children could be used to reduce the number of grid cells of the GeoBlock to query, the number drops with each level, while keeping track of the missing children gets increasingly expensive. 
Therefore, we only consider direct children for this optimization. 
If some of the direct children are cached, we combine their aggregates with the results of the old algorithm for the non-aggregated ones to obtain the final result.

\section{Experimental Evaluation}
\label{sec:evaluation}
We compare GeoBlocks with on-the-fly aggregation approaches on real-world data.
To show that our advantage is not dependent on the indexing strategy, we use different strategies to index the base data of the on-the-fly approaches. 
We also compare GeoBlocks against a pre-aggregating approach, the aR-tree~\cite{DBLP:conf/ssd/PapadiasKZT01, papadias:rtree}. 
However, we do not include the aR-tree in all experiments, as it is designed for rectangular queries and does not directly support polygonal ones.

\subsection{Experimental Setup}
\label{subsec:eval_setup}
\begin{figure}
    \centering
    \includegraphics[width=\linewidth]{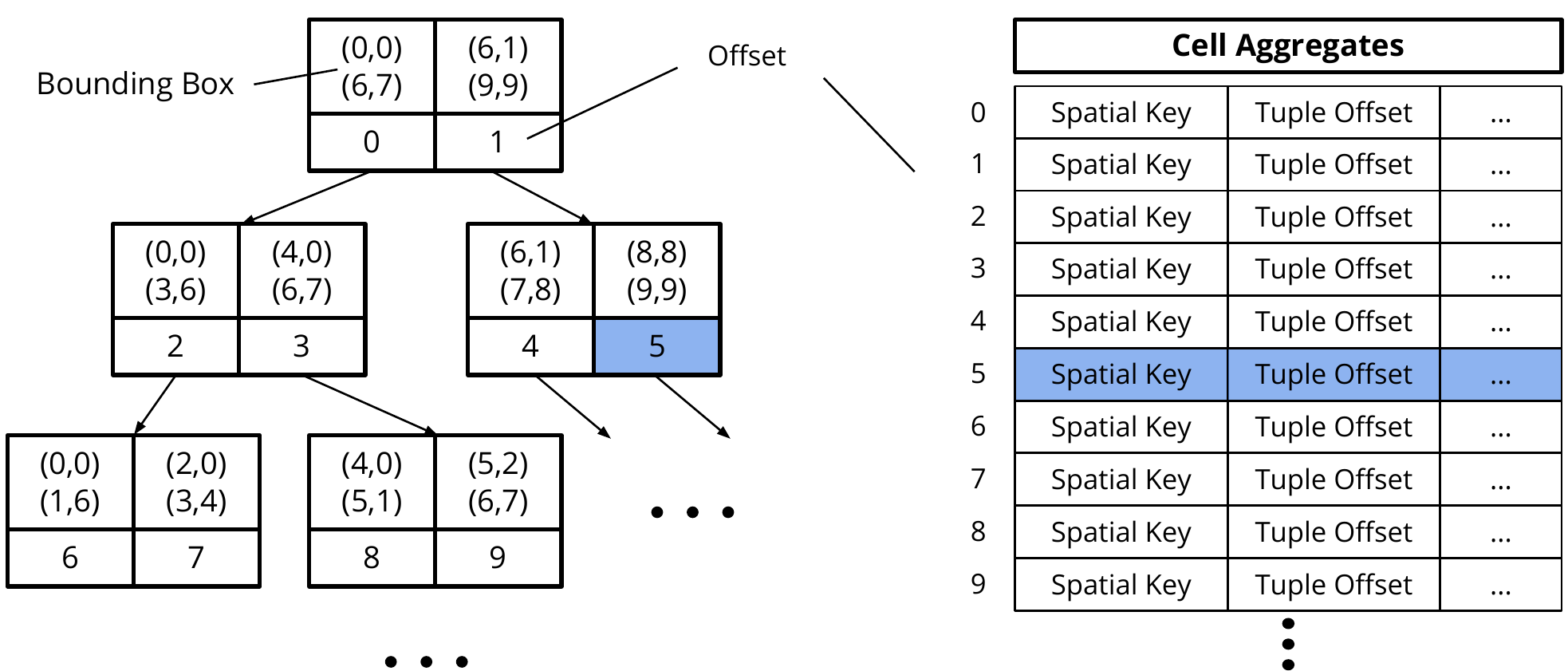}
    \caption{Illustration of aRTree with node size two and offsets into the cell aggregates.}
    \label{fig:artree}
\end{figure}

{\ParHeadSmall Baselines.} To keep the experiments as fair as possible, we use the mapping from geospatial space to linear space for the baselines as an index key unless specified otherwise.
Furthermore, we keep all data in a columnar layout. 
Below, we describe the three strategies that we use to index the raw data, as well as our pre-aggregating baseline:
{\MiniHead BinarySearch:} This is the simplest baseline. Instead of indexing the data, we use binary search to locate the first and last contained raw tuple in the data. Afterward, we loop over all tuples in between and compute the requested aggregates.
GeoBlocks use binary search to locate the cell aggregate in a similar way.
{\MiniHead BTree:} We use the BTree as a secondary index over the raw data.
For the experiments, we use an open-source B-tree implementation by Google~\cite{google:btree}. 
We probe the tree for the first child and scan the sorted raw data until no further tuple qualifies.\footnote{We first tried the PointIndex of the S2 library (\url{https://s2geometry.io/devguide/cpp/quickstart.html\#s2pointindex}) that uses the same b-tree as point storage. 
Initial measurements showed that our optimized BTree implementation outperformed the PointIndex by 3$\times$, so we opted for our implementation.}
{\MiniHead PHTree:} Our last non-aggregating baseline is a multidimensional point index structure, the PH-tree~\cite{zanschke:phtree}. Instead of the one-di\-men\-sio\-nal spatial key, we use the latitude and longitude of the points to index the data. As the PH-tree only supports rectangular range queries, we use S2 to get the interior rectangle of the query polygon and use this as a query region. This way, we hope to keep the comparison fair, if not favorable for the PHTree, as this interior rectangle covers fewer points than our approach. As a consequence, the PHTree's query results differ from the results of the other approaches. For the measurements, we use an open-source C++ implementation~\cite{mcxme:gitphtree}.
{\MiniHead aRTree:} We implement the aR-tree~\cite{papadias:rtree, DBLP:conf/ssd/PapadiasKZT01} based on the boost R-tree~\cite{boost::rtree}. To  minimize overlaps between nodes and thereby optimize the query performance, we use the $R^{*}$ algorithm. In our implementation, each node covers a region $r$ and has up to 16 child nodes, which further subdivide $r$ into smaller areas. 
For each node, we store the aggregates in a cell aggregate corresponding to the region covered by the node, and reference it with an offset (cf. Figure~\ref{fig:artree}).
That way, we can modify the RTree query logic by adding \textit{early abortion} exactly like in the aR-tree. 
Given a search area $s$ and a node of the aR-tree that covers a region $r$, we distinguish three cases, as shown in Listing~\ref{lst:rtreelookup}: (a) If $s$ is completely contained by the covered region $r_c$ of one of $n$'s child nodes, we recursively continue the search at the child node and do not consider other overlapping child nodes as this would result in counting values multiple times. (b) If the region covered by a child node is completely contained within the search area, we add its aggregated value to the overall result and continue processing the next child node. (c) If $s$ and the child node region intersect, we \textit{mark} the child node to be processed later \textit{iff} no other child node fulfills criterion (a). 

By accepting that points are counted multiple times in the case of overlapping internal nodes, our aR-tree implementation follows the query algorithm of the original aR-tree that does not consider overlapping children. While the implementation delivers an upper-bound of the result, it visits the internal nodes in the same way the aR-tree does, thus achieving the same performance.

\begin{listing}[t]
\begin{minted}[linenos,frame=lines,escapeinside=||,fontsize=\footnotesize,numbersep=\mintednumbersep,xleftmargin=4mm, breaklines]{python}
def queryARTree(node, searchArea, result):
  partiallyOverlappingNodes = []
  
  for child in node:
    if child.contains(searchArea):
      return queryARTree(child, searchArea, result)
    if searchArea.contains(child):
      result += child.aggregatedResult
    else if searchArea.intersects(child):
      partiallyOverlappingNodes.append(child)
  
  for child in partiallyOverlappingNodes:
    result += queryARTree(child, searchArea, result)
  return result
\end{minted}
\caption{aR-tree lookup query}
\label{lst:rtreelookup}
\end{listing}

{\ParHeadSmall Implementation.} 
We implement GeoBlocks in C++ as described in Section~\ref{sec:geoblocks}. Our implementation, as well as that of all baselines, is single-threaded. Throughout this section, especially in all figures, we will refer to GeoBlocks as Block. Furthermore, we will differentiate between the regular Block and Block$^\mathrm{QC}$. Block denotes GeoBlocks without query caching using the basic query algorithm. Block$^\mathrm{QC}$ is GeoBlocks using query caching with the AggregateTrie and adapted query process outlined in Figure~\ref{fig:geoblocks_queryflow}. %

{\ParHeadSmall Hardware.} 
All experiments are run on a server machine with two Intel Xeon E5-2680 v4 processors clocked at 2.4\,GHz. The machine is equipped with 256\,GiB of DDR4-2400 RAM. %
All performed experiments fit entirely into main memory.

{\ParHeadSmall Dataset.}
The primary dataset used in the experiments is composed of trip records from 12 million NYC yellow cab rides in the time between January and March 2015, which we cleaned of outliers. It is openly available for download from the NYC Taxi and Limousine Commission (TLC)~\cite{taxi:tlc}. 
It contains data from individual rides like pickup and drop-off location and time, passenger count as well as trip distance.

Unless otherwise specified, the queries consist of polygons representing NYC neighborhoods taken from~\cite{neigh-poly}. As a base workload, we build a query containing each polygon once. For the skewed workload, we select 10\% of neighborhoods uniformly at random and query them multiple times. We select 7 aggregates, requesting each column at least once, as query output.

In addition, we use 8 million geotagged tweets from the contiguous US and query them using polygons representing US states. Finally, we use an extract of 389 million OpenStreetMap (OSM) points in the Americas and query them with polygons representing countries. Both these datasets have randomly generated integer values as payload. For both, we fix the level at 11 (\textasciitilde 7km diagonal). Unless otherwise specified, all experiments are conducted on the primary dataset only.

\subsection{Baseline Comparison}
\label{subsec:eval_base}

\begin{figure}
    \centering
    \includegraphics[width=\linewidth]{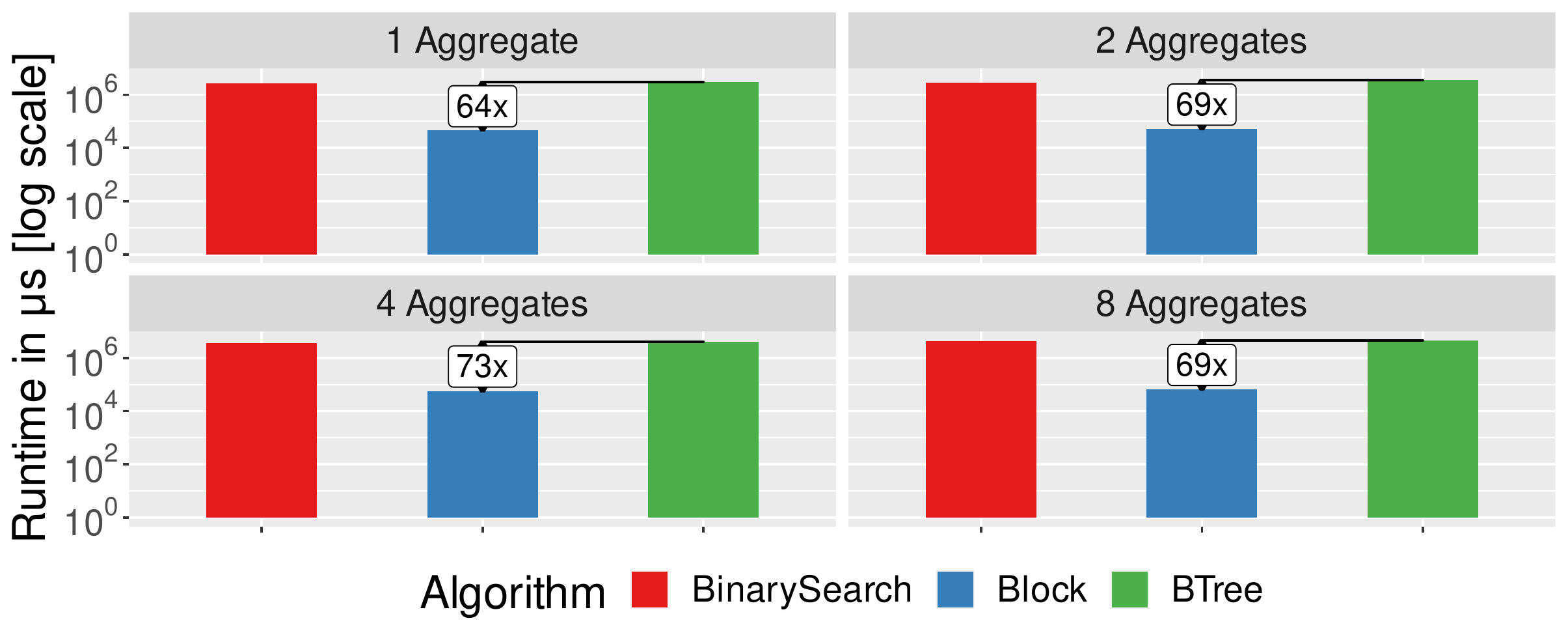}
    \caption{Runtime with increasing number of aggregates.}
    \label{fig:plot_aggs}
\end{figure}

{\ParHead Impact of Number of Aggregates.} To show the impact of the number of aggregates on the performance of the baselines and the Blocks, we use a combined workload consisting of once the base and four times the skewed workload. 
We query this workload for 1, 2, 4, and 8 aggregates and report the results in Figure~\ref{fig:plot_aggs}. 

As one can easily see, GeoBlocks outperform both the BTree and BinarySearch baseline in all cases. 
We omitted the PHTree and aRTree from these experiments, as the imprecise rectangular approximation of the skewed workload lead to a drastic increase in their runtime.
Even for the base workload, the PHTree was slower by a factor of about 3$\times$ while covering fewer tuples.

{\ParHead Indexing Overhead.} 
\begin{figure*}
    \centering
    \begin{subfigure}[t]{0.33\textwidth}
        \centering
        \includegraphics[width=\linewidth]{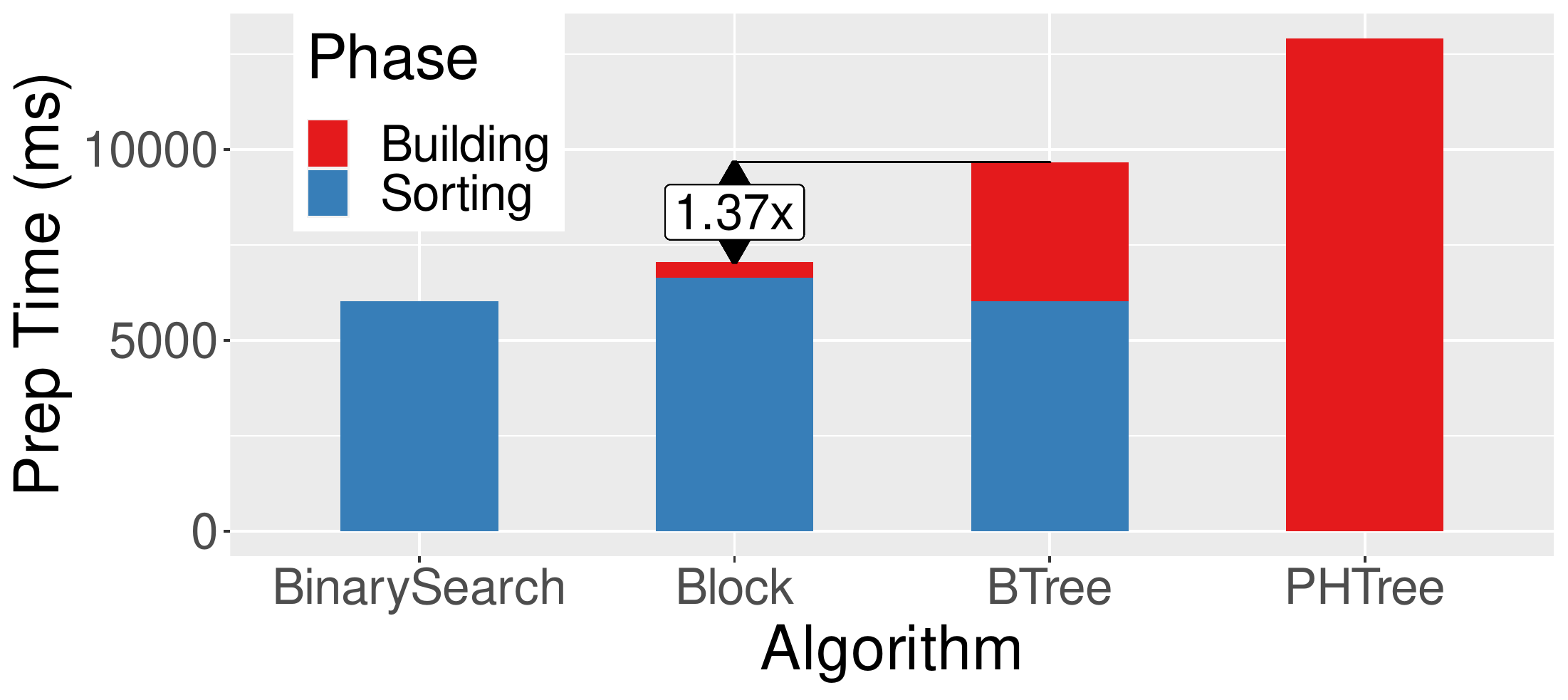}
        \caption{Build time of GeoBlocks and baselines.}   
        \label{fig:plots_timeoverhead}
    \end{subfigure}  
    ~ 
    \begin{subfigure}[t]{0.33\textwidth}
        \centering
        \includegraphics[width=\linewidth]{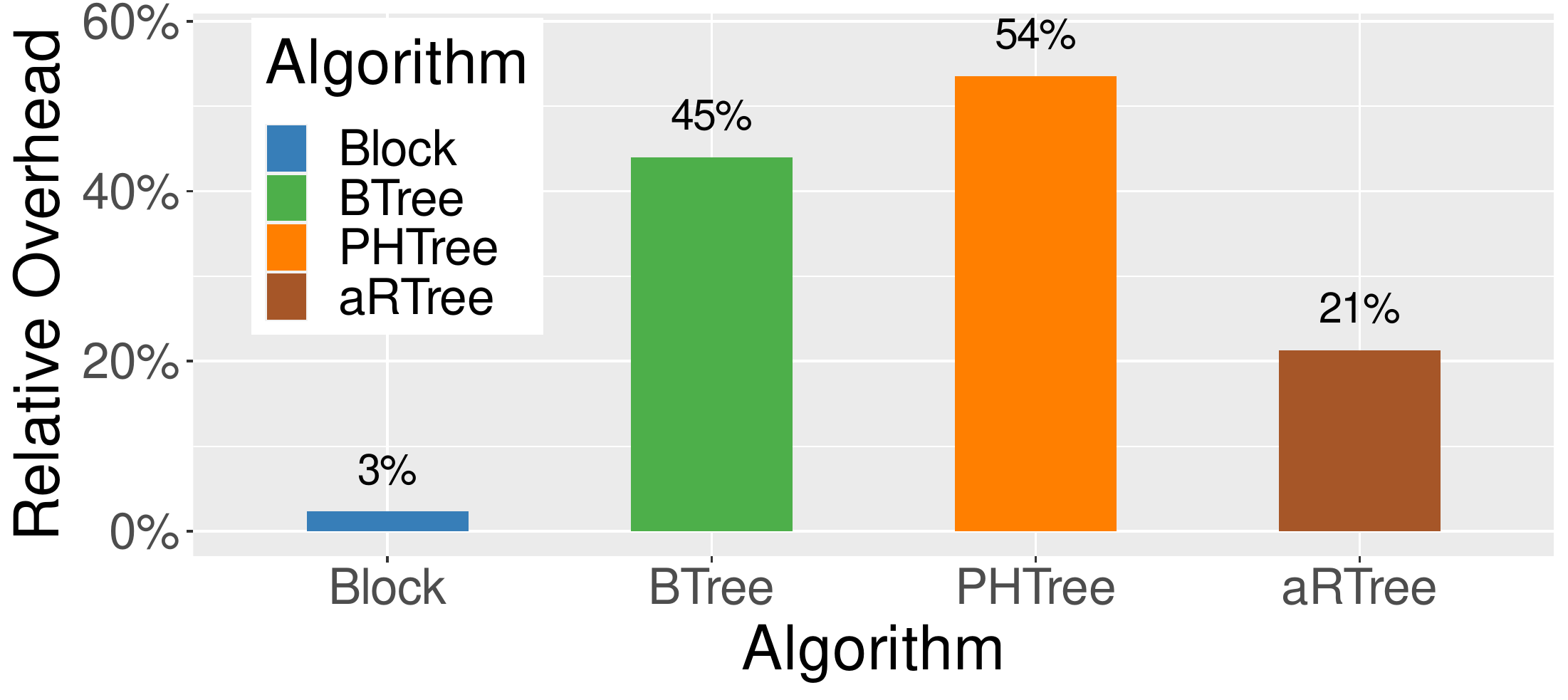}
        \caption{Size overhead of GeoBlocks and baselines.}    
        \label{fig:plots_spaceoverhead}
    \end{subfigure}%
    ~ 
    \begin{subfigure}[t]{0.33\textwidth}
        \centering
        \includegraphics[width=\linewidth]{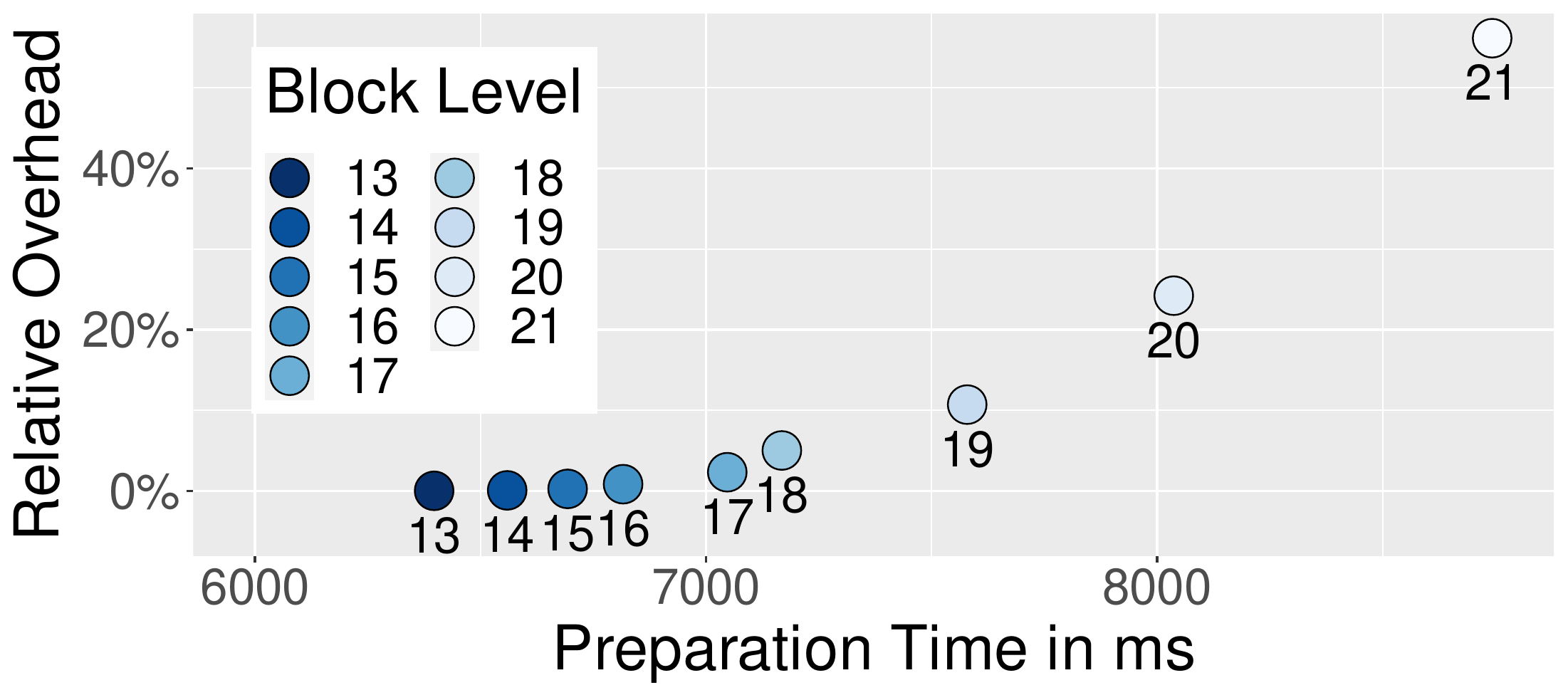}
        \caption{Level influence on GeoBlocks overhead.}   
        \label{fig:plots_levelwiseoverhead}
    \end{subfigure}
    \caption{Index overhead in build time and space.}
    \label{fig:plots_overhead}
\end{figure*}
We compare the build time, \ie the preparation time required prior to running any query, in Figure~\ref{fig:plots_timeoverhead}, with the block level set to 17 (\textasciitilde100m diagonal). 
The reported times for sorting are measured once for the optimized out-of-place sorting for the Blocks and reported for each baseline. This step is completely identical in all sorting baselines. There is a noticeable gap in the sorting phase between the BTree/BinarySearch and the Block. This gap is caused by the collection of grid cell ids to aggregate that we piggybacked on the sorting process to save an additional pass on the data. Overall, the Block is built faster than the BTree and the PHTree, and slightly slower than the BinarySearch, which only needs to sort the input data. We exclude the aRTree baseline from this experiment as we only optimized the implementation for query performance, and the build time was multiple orders of magnitude slower than the others described.
Most notably, the majority of the Block preparation is spent on sorting, indicating that once the data is sorted, building additional Blocks with different filter sets is reasonably cheap.

The relative space overhead of each algorithm is depicted in Figure~\ref{fig:plots_spaceoverhead}. BinarySearch was omitted as it does not require any additional storage. One could argue that this is not a fair comparison to the BTree and PHTree as they index individual points, but as our goal is to provide approximate results, we wanted to show that storing intermediate results is less space-consuming than one would assume for such fine-grained aggregates. While the aRTree is more space-saving when compared to the single-point indices, it still introduces an order of magnitude higher storage overhead than GeoBlocks.

\begin{figure}
    \centering
    \includegraphics[width=\linewidth]{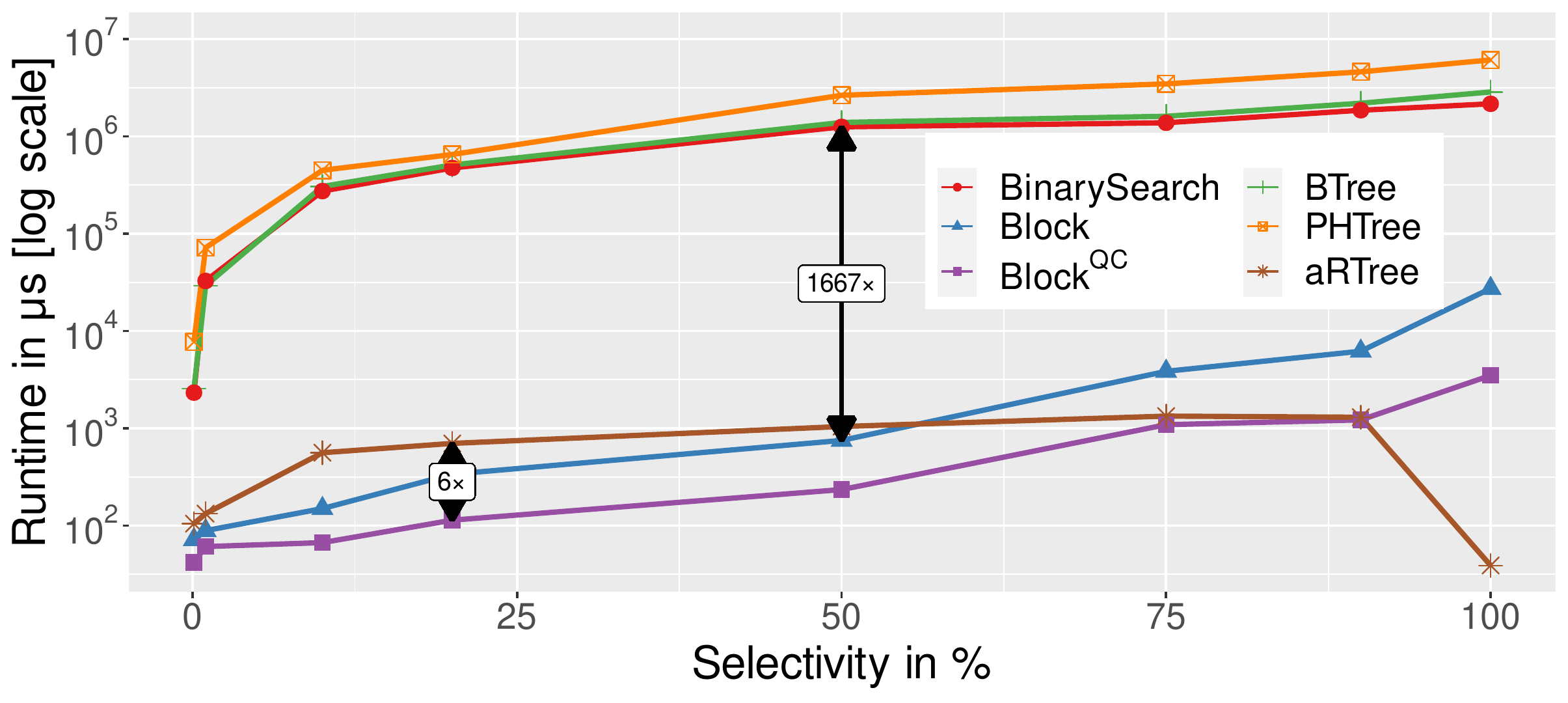}
    \caption{Query runtime for varying selectivity.}
    \label{fig:plots_selectivities}
\end{figure}

{\ParHead Impact of Selectivity.} 
Selectivity is usually defined based on a single query, but in our context, it is hard to specify what a single query is. We break down query polygons, \eg the \textcolor{motorange}{orange} bordered Lower East Side in Figure~\ref{fig:geoblocks_motivational}, to different-sized cells covering the polygon (\eg the \textcolor{motpurple}{purple cell}), which in turn are broken down into equally sized cells to query (\textcolor{motblue}{blue cells}). While the intermediate cells of the query polygon's covering are the best representation of individual queries, as each index is probed once for them, they are artificial concepts introduced by our algorithm. 
Furthermore, these are hard to map to the rectangular query regions of the PHTree and the aRTree. Therefore, we define selectivity based on query polygons. For this experiment, we artificially select polygons covering a part of NYC, which contains a certain percentage of the total rides. Figure~\ref{fig:plots_selectivities} reports the runtime of the base workload at different selectivities using a logarithmic scale.
PHTree's and aRTree's measured selectivities differ slightly from the reported ones due to the less precise covering using an interior rectangle. As this covering contains fewer points, this should slightly skew the experiment in favor of the PHTree and aRTree. 
Even though GeoBlocks can handle rectangular queries as well, since rectangles are just constrained polygons, we opted for the most-precise covering where possible.

While runtime rises quickly for all baselines for selectivities above 1\%, the increase is much softer for both Block variants. Even though the workload is not skewed, and we only use 2\% of additional storage for query caching, Block$^\mathrm{QC}$ still outperforms the non-caching Block across all selectivities. This is likely explained by the shape of the polygons that are often simple quadrilaterals or pentagons. These can be covered using few cells and, therefore, most of these cells can be pre-aggregated. BinarySearch can keep up with the BTree, reporting similar runtimes independent of selectivity, while the PHTree lags behind quickly. Even if the relative runtime gap narrows for higher selectivity, the absolute gap still favors GeoBlocks. The aRTree, our implementation of the aR-tree, outperforms the on-the-fly aggregating benchmarks easily while staying behind GeoBlocks for lower selectivities. However, it can catch up with Block at around 50\% selectivity. At 100\% selectivity, the aRTree needs to only access the root aggregate, explaining the sharp drop in runtime. Overall, GeoBlocks outperform the non-aggregating baselines by at least two and up to three orders of magnitude, performing on-par with the aR-tree while delivering far more precise results.

{\ParHead Scalability.}
\begin{figure}
    \centering
    \begin{subfigure}[t]{0.48\linewidth}
        \centering
        \includegraphics[width=\linewidth]{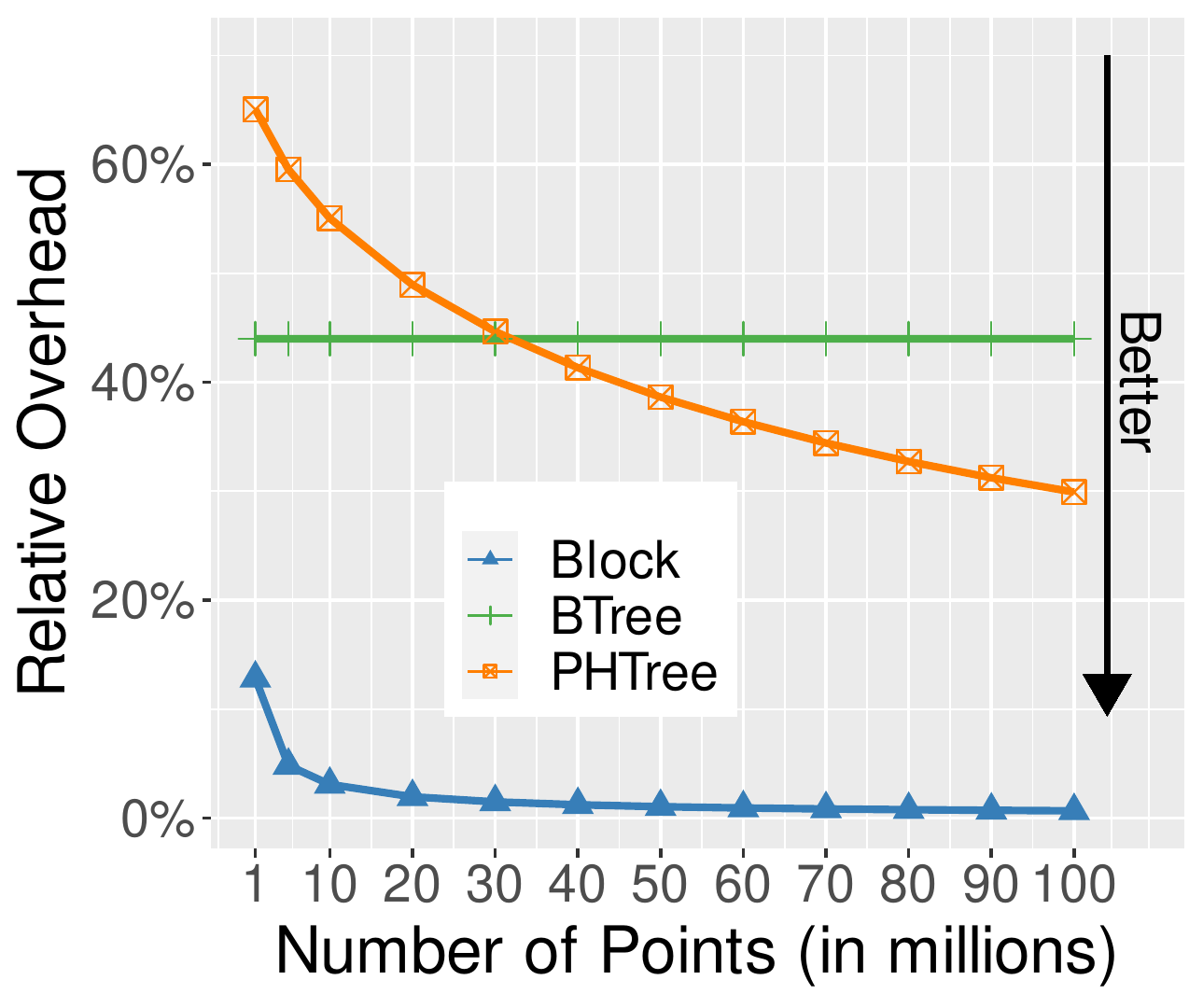}
        \caption{Size overhead of GeoBlocks and baselines.}
        \label{fig:plots_scalesize}
    \end{subfigure}  
    \hfill
    \begin{subfigure}[t]{0.48\linewidth}
        \centering
        \includegraphics[width=\linewidth]{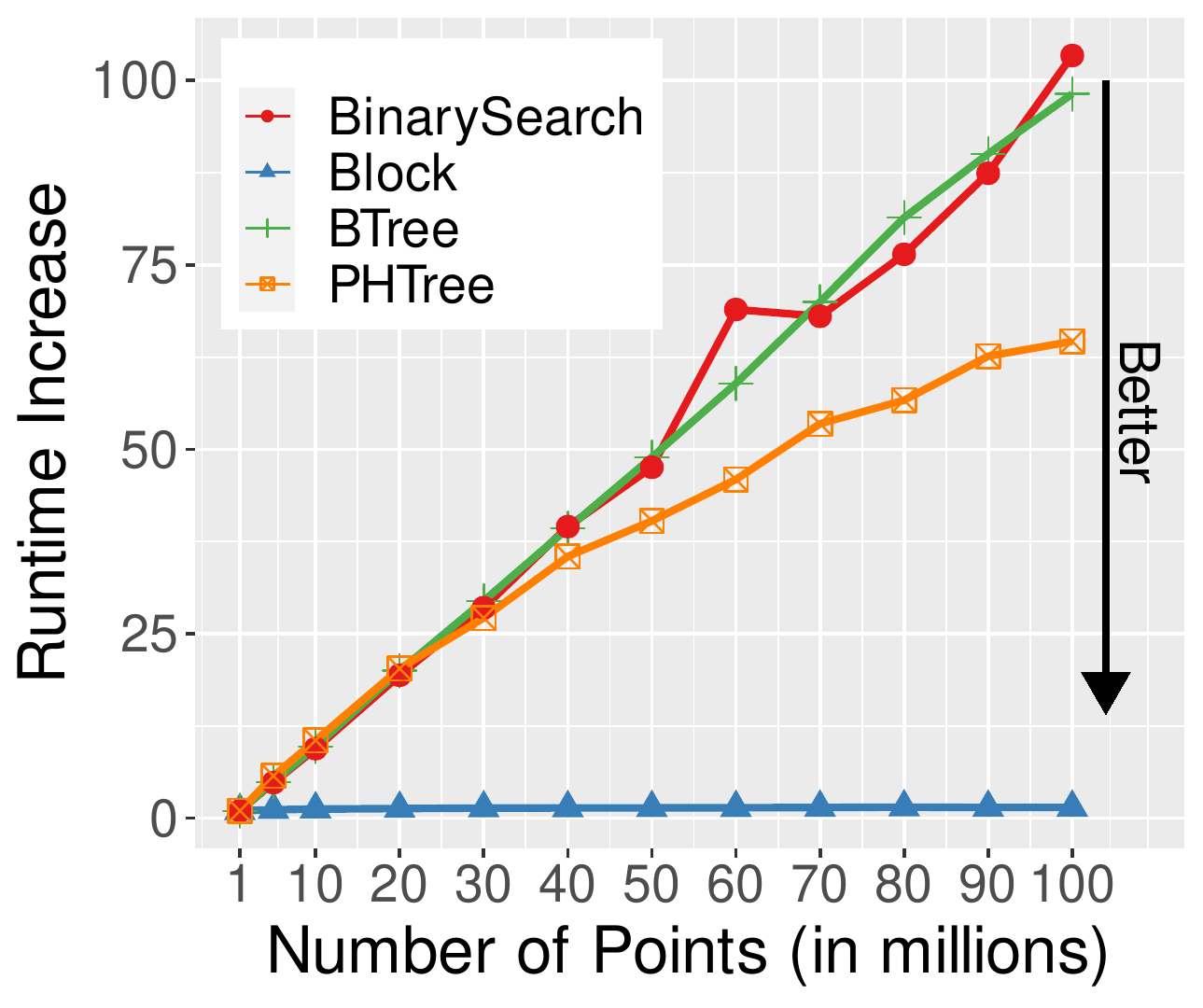}
        \caption{Relative runtime increase of GeoBlocks and competitors compared to 1M points.}   
        \label{fig:plots_scalequery}
    \end{subfigure}%
    \caption{Scaling with increasing input sizes.}
    \label{fig:plots_scale}
\end{figure} 
To study the performance for different-sized datasets, we collect 100M taxi rides spanning all of 2015 and build and query the approaches for an increasing subset of these rides. We omit the aRTree as the build time exceeded reasonable limits upward of 30 million points. As the build time is dominated by the sorting process, which is shared and identical in all approaches, they scale identically in build time. When comparing the size overhead in \Cref{fig:plots_scalesize}, we can see that the BTree overhead is constant as expected. For the PHTree, we see the positive impact of the integrated compression strategies for bigger datasets. Still, the near fixed-size grid aggregates - the size of a GeoBlock is determined by the spatial distribution of points, not their number - enables even smaller overheads for GeoBlocks. To focus on the individual scalability for queries, we analyze the query runtime normalized to the runtime of each approach for one million points.
As shown in \Cref{fig:plots_scalequery}, both the BTree and the BinarySearch scale linearly with the input size, as the on-the-fly aggregation dominates the runtime. 
We expect a similar behavior from the PHTree, but as the covering is less accurate and chosen deliberately smaller, the increase is not fully linear. 
For GeoBlocks, the runtime stays nearly constant, since it depends on the number of maintained aggregates, and not on the number of individual points. 
The number of aggregates is in turn determined by the spatial distribution of the input.
Since one million points already cover most areas in NYC, the distribution does not change when 
further increasing the number of points, \ie the number of aggregates does not increase significantly.
This explains why query latency remains nearly constant for bigger datasets.

{\ParHead Datasets.}
\begin{figure}
    \centering
    \includegraphics[width=\linewidth]{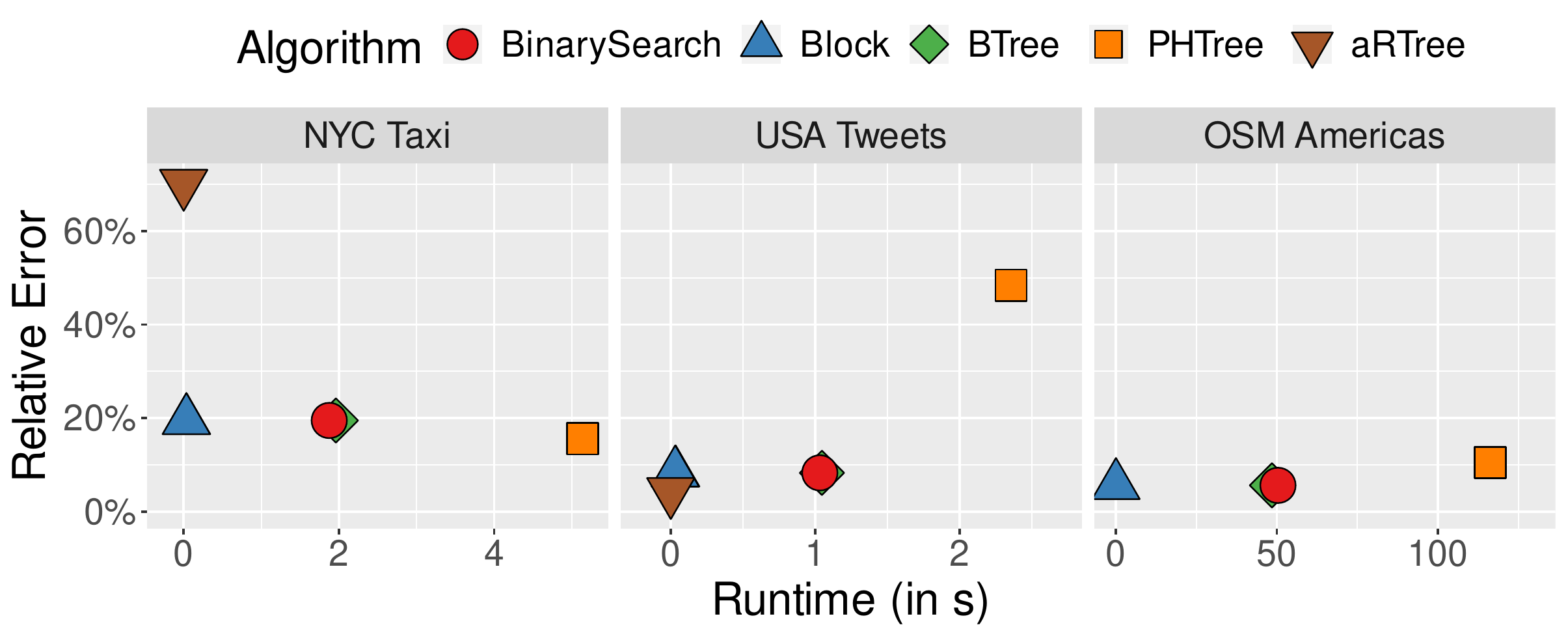}
    \caption{Query runtime and relative error for varying datasets.}
    \label{fig:plot_datasets}
\end{figure}
To show that our approach is not limited to the NYC taxi dataset, we evaluate it on the two additional datasets in \Cref{fig:plot_datasets}. We again query the whole area represented by the individual polygons and report runtime, as well as the average error defined as $\frac{\lvert \textrm{\# tuples in query result} - \textrm{\# tuples in polygon} \rvert}{\textrm{\# tuples in polygon}}$. For the OSM dataset, the aRTree again was excluded because of its excessive build time. As the Block, BinarySearch, and BTree use the same covering, the result and error are identical. While the aRtree and PHTree use an identical rectangular representation, the pre-aggregated nodes of the aRTree lead to a different result, and therefore error. Overall, the aRTree and Block are similarly fast with a slight advantage for the aRTree, outperforming the non-aggregating approaches easily. However, the error for Block is far more stable.

{\ParHead Accuracy.}
\begin{figure}
    \centering
    \includegraphics[width=\linewidth]{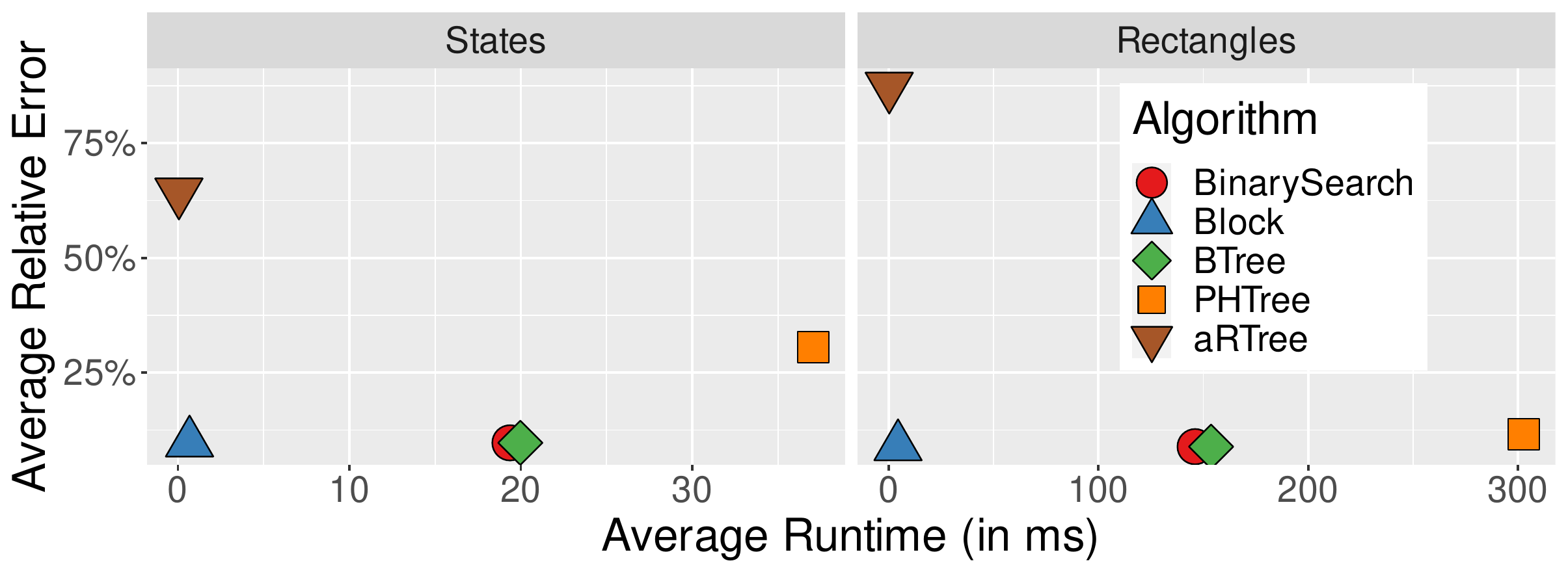}
    \caption{Query runtime and relative error for US states and generated rectangles on the Twitter dataset.}
    \label{fig:plot_rect}
\end{figure} 
Finally, we want to study the influence of smaller individual polygons, as well as rectangular areas, on both runtime and relative error. Therefore, we query all US states and 51 randomly generated rectangles within the US on the Twitter dataset and report the average runtime and error in \Cref{fig:plot_rect}. In contrast to the previous experiment, we query all areas individually. For both polygons and rectangles, the same overall trends are visible. The aRTree is slightly faster than Blocks as the large polygons can be answered in the upper levels of the tree. However, this leads to high imprecision even for rectangular queries as partially overlapping internal nodes might be counted multiple times. Besides, we see that the individual errors canceled out in \Cref{fig:plot_datasets}, leading to a seemingly good error bound. While the PHTree error also improves considerably for the rectangular workload, we expected it to be exact. We suspect this is caused by our transformation of the coordinates to integer space, which is necessary for efficient queries. As expected, the performance of Blocks and the other approximating baselines does not degrade for rectangular areas. The aggregating approaches again far outperform the point indexing approaches in runtime.

\subsection{Sensitivity Analysis}
\label{subsec:eval_config}

After showing that GeoBlocks easily outperform all baselines, we study the impact that the configuration of GeoBlocks has on throughput, as well as the impact of data skew on the adaptive Block version. 
The Block configuration is specified by three parameters: 
The first setting we study is the level of the Block, \ie the resolution of the grid overlying the spatial domain. 
Next, we take a look at the impact of skew on both Block and Block$^\mathrm{QC}$. Finally, we examine how the size of the AggregateTrie influences the runtime of unskewed and skewed workloads.

\begin{figure}
    \centering
    \includegraphics[width=\linewidth]{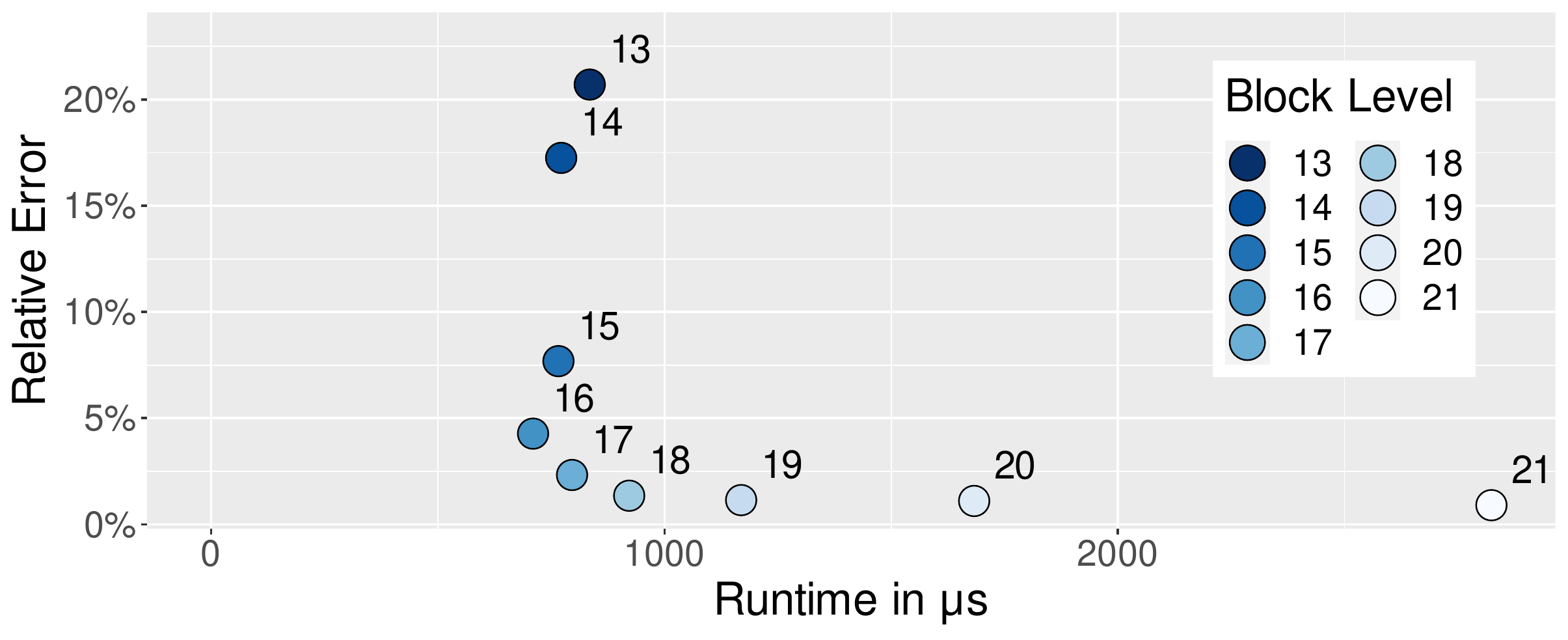}
    \caption{Relative error and runtime at varying levels.}
    \label{fig:plots_error}
\end{figure}

{\ParHead Impact of Block Level.}
We vary the block levels from 13 to 21 (between \textasciitilde1.5km and \textasciitilde6m diagonal) while keeping the other configuration parameters fixed. 
From a runtime-only point of view, lower-level (coarser-grained) blocks are always preferable, as the query algorithm needs to take fewer cells into account. 
However, this comes at the price of precision loss. 
Figure~\ref{fig:plots_error} illustrates the connection between the block level, the runtime, and the relative error introduced by the cell covering.
The cell covering can introduce only false positive results, \ie some reported results are not contained in the actual polygon.
The figure clearly shows the expected overall trend: the higher the level, the lower the relative error and the higher the runtime. 
However, after a certain point, decreasing the level further does not pay off. Further, we see that the correlation between error and runtime is not linear, as we already suspected in \Cref{subsec:geo_error}. The correlation does not even follow the discussed influences completely, which is likely caused by missing sparse children, and the non-uniform distribution of points leading to a gap between the relative error and the configurable spatial error.

\begin{table}
    \centering
    \footnotesize
\begin{tabular}{@{}lrrrrrrrrr@{}}
\toprule
\textbf{Level}\hspace{-1mm}    & 13 & 14 & 15 & 16 & 17 & 18 & 19 & 20 & 21 \\ \midrule
\textbf{Sorting}\hspace{-1mm}  & 6020   & 6008  & 6317   & 6459   & 6633   & 6754   & 7028   & 7344   & 7666   \\
\textbf{Building}\hspace{-1mm} & 376   &  499  &  376  & 356   &  411  &  408  &  538  &   666 &  1025  \\ \bottomrule
\end{tabular}
    \caption{Index build times in ms at varying levels.}
    \label{tab:eval_times}
\end{table}

The block level influences not only the relative error and the runtime, but also the build time and size of GeoBlocks. Figure~\ref{fig:plots_levelwiseoverhead} depicts the build time and size overhead for GeoBlocks from levels 13 to 21. The build time seems to be only slightly affected by the level, rising slowly with it.
Table~\ref{tab:eval_times} splits the runtime into two parts: sorting and building.
There is a noticeable increase in sort time along with the block level, in addition to the expected increase in build time. This increase in sorting can be explained through our grid cell extraction that we piggybacked to the sorting process, which has to extract more finer-grained cells. 
The size overhead, however, grows exponentially due to the exponentially growing number of cells along with the level. 

{\ParHead Impact of Skew.}
\begin{figure}
    \centering
    \includegraphics[width=\linewidth]{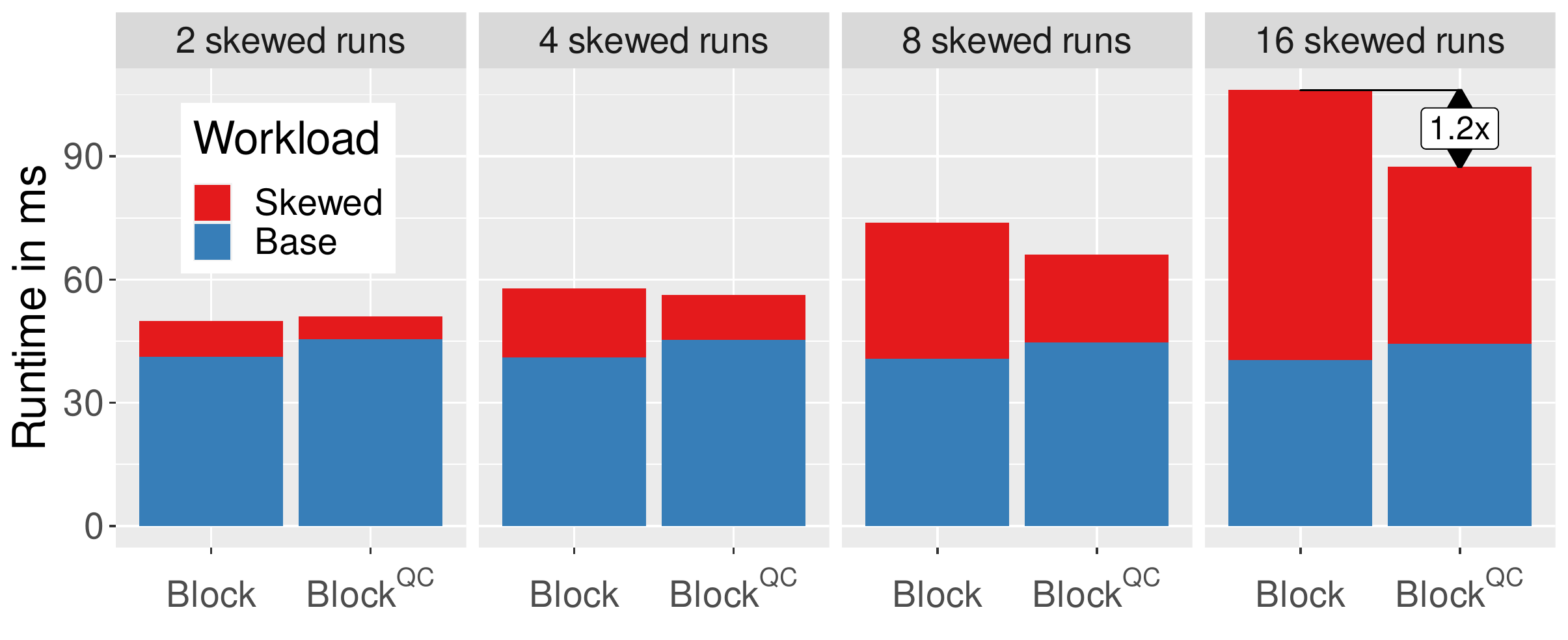}
    \caption{Query runtime with increasing workload skew.}
    \label{fig:plots_skew}
\end{figure} 
To study the impact of data skew on the effectiveness of query caching, we measure the query runtime when running the NYC workload once, and the skewed workload multiple times.
The number of times we run the skewed workload varies in each experiment.
We fix the block level to 17 (\textasciitilde100m diagonal) and the size of the cache to 5\% of the cell aggregates, which roughly corresponds to aggregating all cells of the skewed workload. 
Figure~\ref{fig:plots_skew} displays the absolute runtime for both the base and the skewed part of the workload. 
One can see that after four skewed runs, the cached aggregates start to pay off. 
With even more skew in the total workload, our query-caching Block$^\mathrm{QC}$ quickly starts to outperform Block. 
Furthermore, as expected, the runtime for the base workload stays nearly constant, and is always slightly faster for Block. 
This is easily explained by the overhead of probing the AggregateTrie for each cell, regardless of whether the cell is aggregated or not. 

\begin{figure}
    \centering
    \includegraphics[width=\linewidth]{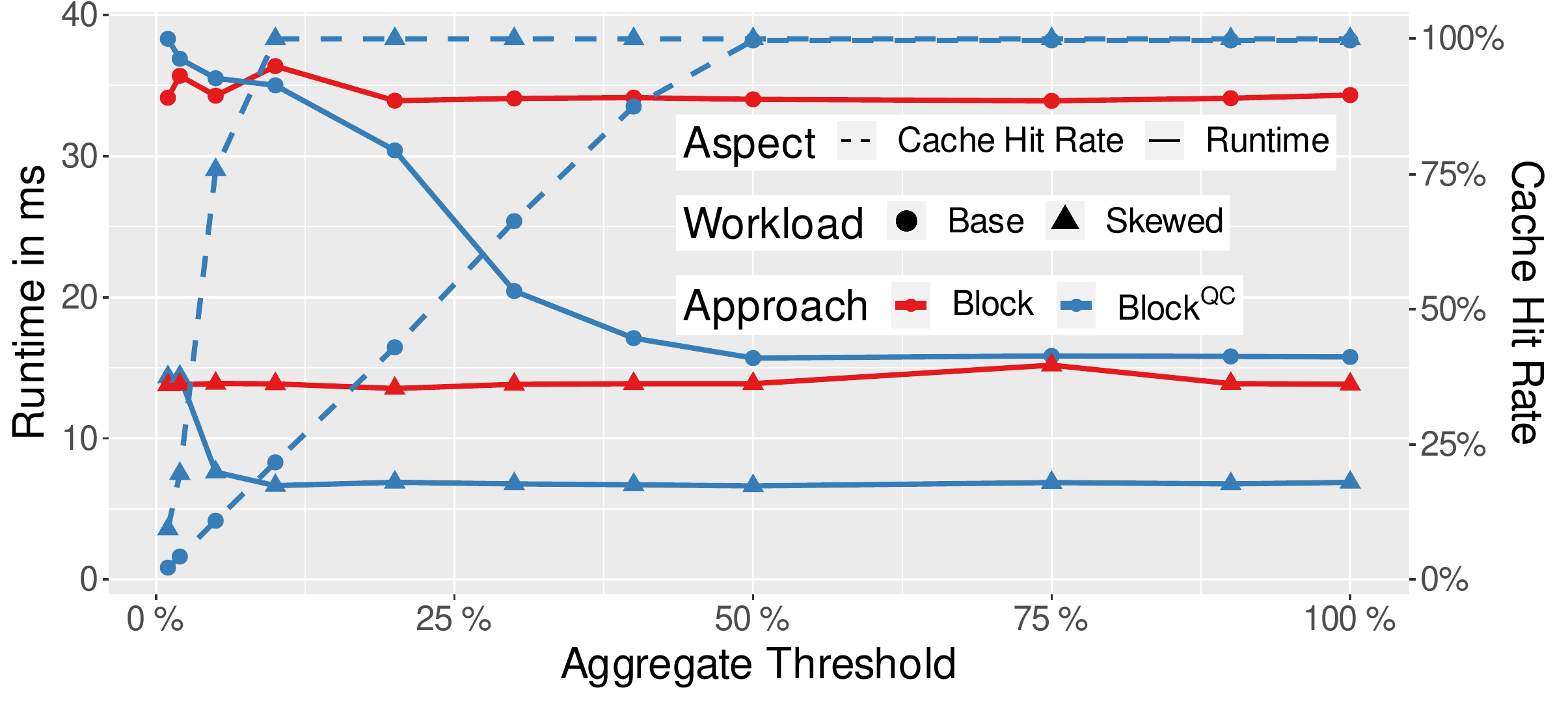}
    \caption{Impact of threshold on workload runtime (solid line) and cache hit rate (dashed line).}
    \label{fig:plots_thresh}
\end{figure}

{\ParHead Impact of Aggregate Threshold.}
Having studied the impact of skew, we want to examine how the aggregate threshold, and thereby the size of the query cache (in Block$^\mathrm{QC}$), influences the runtime of the base and the skewed workload. 
The aggregate threshold denotes the relative size overhead that the query cache, the AggregateTrie, introduces compared to the size of the cell aggregates in the regular GeoBlock. 
We again fix the block level to 17, and the number of skewed runs to four.
Figure~\ref{fig:plots_thresh} depicts the measured runtimes and cache hit rates. The runtime of Block is unaffected by the changed threshold and only acts as a baseline to highlight the influence on Block$^\mathrm{QC}$. Up until a threshold of around 5\%, only queries from the skewed workload can be answered using the AggregateTrie. The small speedup in the base workload can be explained by the inclusion of the skewed workload in the base workload. Once all cells in the skewed workload are cached, and the cache hit rate for the skewed part reaches 100\%, other query cells of the base workload start to get cached as well. While this, of course, leads to further runtime improvements, it is undesirable, especially when memory is scarce. 
In our experiments, at around 50\%, the cache hit rate reaches 100\% for both workloads, and there is no further speedup, even when the cache size is doubled. The cache hit rate, illustrated by the dashed line and shown on the right axis, shows the desired effect. The skewed part is cached almost immediately, and the hit rate for the unskewed workload grows linear with increasing cache size. The average lookup time slowly grows from 58ns at 1\% to 81ns at 100\%. As the lookup time depends on the number of levels (30 in the maximum) and not on the size, this growth is attributable to more complex access patterns for larger cache trees.

\subsection{Changing Filters}
\label{subsec:eval_multi}

\begin{figure}
    \centering
    \includegraphics[width=\linewidth]{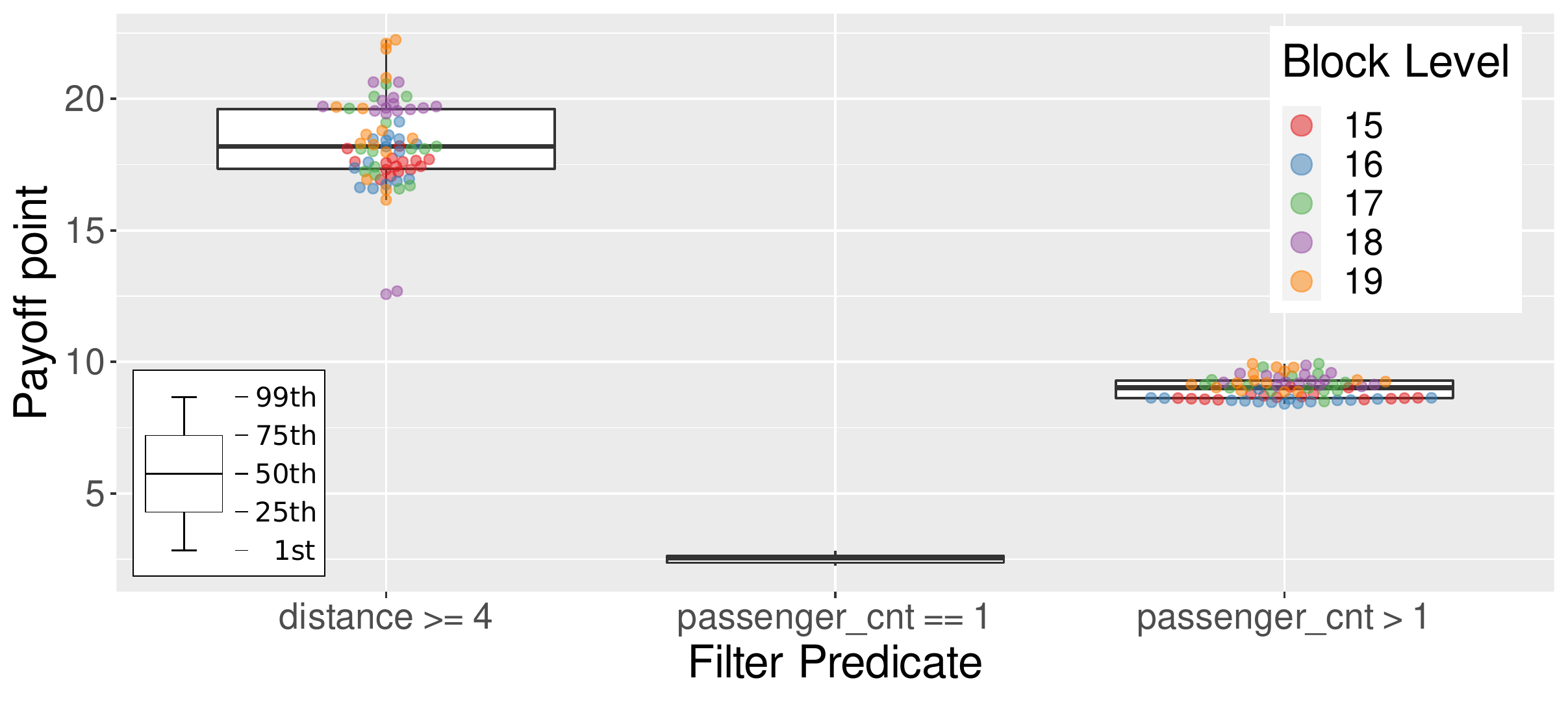}
    \caption{Payoff point: Number of incremental builds required to amortize the cost of sorting the raw data.}
    \label{fig:plots_multipay}
\end{figure}
Finally, we compare our process of \Cref{fig:process}, wherein we build multiple GeoBlocks from the sorted base data, against building isolated GeoBlocks from scratch. 
We vary the block level from 15 to 19 (between \textasciitilde420m and \textasciitilde27m diagonal), and build 15 GeoBlocks per level using three different predicates of varying selectivity:
\begin{itemize}
    \item \textbf{distance >= 4:} Long taxi trips, selectivity of \textasciitilde16\%    
    \item \textbf{passenger\_cnt == 1:} Solo taxi trips, selectivity of \textasciitilde70\%
    \item \textbf{passenger\_cnt > 1:} Shared taxi trips, selectivity of \textasciitilde30\%
\end{itemize}

For this, we want to analyze how many different filter and level combinations are required to amortize the initial cost of sorting. 
Figure~\ref{fig:plots_multipay} shows the payoff point of filter changes for our three filter predicates. 
The payoff point is the number of incremental builds required to be, in sum, faster by creating incremental builds than building individual GeoBlocks from the raw data and filtering before sorting. 
We omitted the individual runtimes for the \texttt{passenger\_cnt == 1} predicate as they would be too densely packed vertically.

As expected, the more selective the filter, the lower the speedup. Once all tuples in the raw data have been filtered according to the predicate, the qualifying tuples have to be sorted. 
More selective predicates take longer to amortize as sorting few tuples is cheap
, whereas for the 70\% selectivity query \texttt{passenger\_cnt == 1}, the more expensive sorting is amortized almost immediately. 
There is a correlation between the block level and amortization, most notably for the most selective predicate \texttt{distance >= 4}. 
Given that the payoff point drastically rises with lower selectivity, we expect that incremental builds will only pay off when the new filters are less selective.
If only a few highly selective queries are expected, building regular GeoBlocks directly from the raw data will still be the fastest option. %
However, the time to switch to a new filter, and therefore the individual query latency, will always be lower for incremental builds.

\section{Discussion}
\label{sec:discussion}
In this section, we discuss the takeaways of the evaluation as well as updates for GeoBlocks.%

{\ParHeadSmall Evaluation Summary.} First, we showed that pre-aggregation in a spatial context pays off when a limited and bounded spatial error is acceptable, independently of the number of aggregates queried and the selectivity of the query polygons. 
Furthermore, GeoBlocks can be built fast, introducing only a small overhead compared to the simple BinarySearch baseline.
Even when the data is already indexed with one of our baselines
 (\ie without taking the index build time into account),
GeoBlock's build time of around 7 seconds can be amortized by fewer than 30 polygon queries with a selectivity of 10\% (\cf Figures~\ref{fig:plots_timeoverhead}~and~\ref{fig:plots_selectivities}). %
In addition, building multiple GeoBlocks once the data is sorted is possible within one second for our dataset, \cf Figure~\ref{fig:plots_timeoverhead}.  
Building new GeoBlocks for different filters is even faster when using sorted base data, often amortizing the initial extra cost of sorting all data in less than 10 filter changes (\cf Figure~\ref{fig:plots_multipay}). 
Even though not all configurations are optimal for GeoBlocks, there are acceptable error-runtime trade-offs, in our case around levels 17 and 18. 
While the level does not significantly impact the index build time, the size overhead growth is almost exponential, \cf Figure~\ref{fig:plots_levelwiseoverhead}, indicating that it is wise to think about which error is acceptable for the given query workload when memory is scarce.

{\ParHeadSmall Updates.} Up until now, we considered GeoBlocks to be read-only as they are designed for historical point data. However, the layout of GeoBlocks allows us to integrate updates easily, as long as a cell aggregate for the region of the newly arriving tuple already exists. For the non-adaptive version, all we have to do is locate the cell aggregate containing the tuple and update all stored aggregates. 
In the adaptive version, we additionally need to update all cached parents of the grid cell in the AggregateTrie as well. Thanks to the prefix-based indexing property of the trie, we can do this in a single depth-first traversal. Only if tuples arrive for a new, previously unaggregated region, we have to rebuild the aggregate layout, as we rely on the cell aggregates to be sorted. 
However, as we have shown in the evaluation, recalculating the cell aggregates is often possible within a second, so this operation would not induce too much delay when updates are implemented in batches instead of single tuples. 
Other indexing approaches on the cell aggregates (\eg a clustered B-tree) could eliminate the need to rebuild by reserving storage for new aggregates. Preliminary experiments using \texttt{std::map} and a B-tree as an index showed similar lookup performance at the cost of increased size overhead.

\section{Related Work}
\label{sec:relatedwork}
Our approach builds on seminal work from decades of research on spatial indexing.
Decomposing space into hierarchical grid cells~\cite{finke:quadtree, DBLP:journals/csur/Samet84, DBLP:conf/pods/ArefS90}, as well as approximating polygons using simpler shapes~\cite{DBLP:conf/ssd/KriegelHS91}, are all well-known approaches.
Likewise, enumerating cells using a space-filling curve such as Hilbert or Z order~\cite{orenstein:zorder, DBLP:conf/sigmod/Orenstein86} and storing aggregate information within cells~\cite{DBLP:conf/ssd/PapadiasKZT01, DBLP:reference/gis/LazaridisM08a, DBLP:journals/pvldb/SinglaEAM19} are ideas that have been around for some time.
However, while building on these established concepts, GeoBlocks present the first pre-aggregating data structure that supports a \textit{bounded, distance-based error} on the results of \textit{polygonal queries}.
Specifically, prior work on pre-aggregation~\cite{han:spatial, DBLP:conf/ssd/PedersenT01, DBLP:conf/ssd/PapadiasKZT01, papadias:rtree} is limited to rectangular queries and requires an expensive post-processing (refinement) step to answer polygonal queries.
GeoBlocks, on the other hand, yield error-bounded results and do not require expensive refinement.

{\ParHeadSmall Spatial Aggregation.}
Past work has proposed several approach\-es for spatial aggregation queries~\cite{lopez:spatialsurvey}.
These approaches mainly rely on pre-aggregation~\cite{han:spatial,DBLP:conf/ssd/PedersenT01}: they pre-aggregate records at various spatial resolutions and store this summarized information in a hierarchy of rectangular regions, maintained using a spatial index like the quadtree or the R-tree~\cite{lazaridis:MRT, papadias:rtree, DBLP:conf/ssd/PapadiasKZT01, papadias:movingagg}.
For instance, the aRtree~\cite{DBLP:conf/ssd/PapadiasKZT01,papadias:rtree} enhances the R-tree by storing aggregate information for each node.
This allows to directly extract the aggregate of all the records contained in a node, if the node's MBR is fully enclosed in the query region. 
Being a variant of the R-tree, the aRtree constrains the supported queries to only rectangular regions. 
Furthermore, the computed aggregates are approximate and the error \emph{cannot be bounded}, since the accuracy depends on the resolution of the rectangular R-tree nodes.
Providing precision guarantees for arbitrary polygons requires accessing the raw data and involves additional processing.
There are also approaches that store aggregates inside a data cube~\cite{braz:trajectorywarehouse, rao:hierarchy}, or using sketches~\cite{tao:sketches}. 
Nano\-cubes~\cite{lins:nanocubes}, for example, store the CUBE operator for spatio-temporal datasets, and are specifically designed 
for visualization systems.
The data cube-based approaches suffer from the same limitations as the aRtree, since they also rely on a hierarchy of rectangular regions.
Besides, accessing the raw data to refine the aggregates might require additional indices, as the cube does not store individual records. 
Vorona et al.~\cite{deepspace} approximate the distribution of geospatial points with an autoregressive deep learning model to answer arbitrary polygonal queries, but they cannot provide any error bounds.
Pandey et al.~\cite{learnedspatial} propose to use learned indices for query-efficient spatial indexing, albeit limited to range queries.
Finally, Raster Join~\cite{DBLP:journals/pvldb/ZacharatouDASF17} uses GPU rendering to compute aggregates over a point-polygon join.
In contrast, GeoBlocks support aggregation over spatial selections.

Prefix sums~\cite{DBLP:conf/icde/GeffnerAAS99} can be used in addition to pre-aggregation to enable fast range-sums.
This is achieved by only inspecting the aggregates in the two corners of a query region, rather than every aggregate inside the query region.
An example of this is our \inlinesql{COUNT} algorithm. 
However, in contrast to our \inlinesql{SELECT} queries, these range-sums are unable to extract min and max aggregates.

{\ParHeadSmall Materialized Views and OLAP Cubes.}
GeoBlocks are essentially materialized views over geospatial data with support for filters and aggregations. 
In contrast to regular views~\cite{shmueli:matviews, gupta:matviews}, GeoBlocks are designed for historical spatial data and can adapt to the query workload at a fine-grained level using a trie-like cache. 
Work on materialized view selection~\cite{baralis:matviewselection} also makes materialization decisions based on the query workload, but at a much coarser granularity (\eg what columns to aggregate).
There has also been a lot of work on data cubes and query caching \cite{gray:cubes, harinarayan:efficientcubes, shim:caching}, but these do not support geospatial data as a first-class citizen.

{\ParHeadSmall Spatial Point Indexing.} Spatial point indexing approaches typically index points using a hierarchy of MBRs, most notably the R-tree~\cite{guttman:rtree}, or by subdividing grid cells into equally-sized children, \eg the quadtree~\cite{finke:quadtree, DBLP:journals/csur/Samet84}.
Both of these index structures are queried using the dimension-wise min/max values, \ie the query regions are rectangular.
Other approaches, like the UB-tree~\cite{bayer:ubtree}, assign univariate keys to the indexed regions first and rely on these keys for data access.
While the UB-tree does not specify how these keys have to generated, most approaches use space-filling curves like the Z order~\cite{orenstein:zorder, DBLP:conf/sigmod/Orenstein86}.

Based on these concepts, more specialized indices have been developed. The PH-tree~\cite{zanschke:phtree} combines a quadtree with hypercubes to allow splitting all dimensions in each node, providing a space-efficient index structure for multidimensional data.
The space efficiency can be partly attributed to the utilization of prefix sharing, similar to the one used in our trie-like cache. 
Alternating the indexed dimensions in an in-memory tree structure, the BB-tree~\cite{DBLP:conf/icde/0001SL19} offers fast point and range queries for multidimensional data.
While these structures require the index to be built a priori, there are others like QUASII~\cite{pavlovic:quasii}, where the index is built incrementally as a side product of query execution.
As a result, QUASII can adapt to the query workload at runtime. %
However, QUASII only supports spatial range (window) queries.
Recently, Shin et al.~\cite{DBLP:conf/gis/ShinMA19} proposed integrating grid indices into a tree structure to achieve faster node accesses and point operations.

\section{Conclusions}
\label{sec:conclusion}
We have introduced GeoBlocks, a novel pre-aggregating data structure for geospatial data. 
GeoBlocks pre-compute aggregates over fine-grained grid cells, thereby supporting arbitrarily shaped polygons.  
Using these aggregates, GeoBlocks can provide fast query results with a user-controlled spatial error. 
Furthermore, GeoBlocks can speed up aggregate queries for commonly queried regions by dynamically adapting to any given workload using limited additional storage.

Comparing GeoBlocks with on-the-fly aggregating indexing baselines, we have shown that we can outperform them for any number of aggregates, in parts by three orders of magnitude.
The introduced storage overhead is comparable, and often even lower, to that of traditional indexing structures, while GeoBlocks can be built equally fast.
Looking at  GeoBlocks' configuration options, we have shown how they can be adapted to the given dataset and workload, and how they influence the runtime, the overhead, and the error in the result.
Overall, GeoBlocks are materialized views over geospatial data that support filter predicates and aggregates while enabling fine-grained adaptation to the query workload. %

\bibliographystyle{abbrv}
\bibliography{main}  

\end{document}